\documentclass[twocolumn]{aastex631}

\usepackage{amsmath,mathtools,mathrsfs}
\usepackage[T1]{fontenc}

\usepackage{hyperref}

\usepackage{natbib}
\usepackage{comment}
\hypersetup{
  colorlinks=true,        
  linkcolor=blue,        
  citecolor=blue,   
}

\begin{document}

\title{Characterizing Power Spectra of Density Fluctuations in GRMHD Simulations of Black Hole Accretion Using Taylor's Frozen-in Hypothesis}
\author[0000-0001-8757-9323]{Pravita~Hallur}
\affiliation{GRAPPA and Institute of High-Energy Physics, University of Amsterdam, Science Park 904,
1098 XH Amsterdam, The Netherlands}
\affiliation{Indian Institute of Science Education and Research Mohali. Sector 81, SAS Nagar, Mohali, PO Manauli, Punjab 140306, India}
\affiliation{School of Natural Sciences, Institute for Advanced Study, 1 Einstein Drive, Princeton, NJ 08540, USA}
\author[0000-0003-2342-6728]{Lia~Medeiros}
\affiliation{Center for Gravitation, Cosmology and Astrophysics, Department of Physics, University of Wisconsin–Milwaukee, P.O. Box 413, Milwaukee, WI 53201, USA}
\affiliation{Department of Astrophysical Sciences, Peyton Hall, Princeton University, Princeton, NJ, 08544, USA}
\affiliation{School of Natural Sciences, Institute for Advanced Study, 1 Einstein Drive, Princeton, NJ 08540, USA}
\author{Pierre Christian}
\affiliation{WattTime, 490 43rd Street, Unit 221, Oakland, CA 94609, USA}
\author[0000-0001-6952-2147]{George~N.~Wong}
\affiliation{Princeton Gravity Initiative, Princeton University, Princeton NJ 08544, USA}
\affiliation{School of Natural Sciences, Institute for Advanced Study, 1 Einstein Drive, Princeton, NJ 08540, USA}

\begin{abstract}
We characterize the spatial power spectrum of density fluctuations in magnetohydrodynamic flows in a suite of high-resolution, long-time-span general relativistic magnetohydrodynamic (GRMHD) simulations. Extracting the local spatial power spectrum in curved spacetime directly from GRMHD simulations can be challenging for several conceptual and mechanical reasons, including choices of the reference frame, the non-uniform co-ordinate grid of the outputs and limited resolution. Taylor's frozen-in hypothesis describes a mapping between the temporal and spatial power spectrum of turbulence, which we apply to density fluctuations.
We explore the validity of the assumptions underlying Taylor's hypothesis and evaluate its applicability in extracting spatial power spectra of density fluctuations of black hole accretion flows.
Using outputs from the GRMHD code {\tt KORAL}, we explore models with strong and ordered magnetic fields (MAD, Magnetically Arrested Disks) as well as weak and disordered magnetic fields (SANE, Standard and Normal Evolution). We explore the effects of black hole spin on the power spectra and characterize their spectral properties as a function of distance from the black hole. The observed power spectra follow a broken power law with two slopes separated by a break frequency. Our analysis shows a decrease in break frequency with increasing radius, with distinct trends between SANE and MAD flows. We also observe the first slope to be steeper for SANE flows and some notable distinctions between prograde and retrograde spins. 

\end{abstract}

\section{Introduction}\label{sec:1}
High-resolution images of the Galactic Center black hole Sagittarius A$^*$ (Sgr A$^*$, \citealt{2022ApJ...930L..12E, 2022ApJ...930L..13E, 2022ApJ...930L..14E, 2022ApJ...930L..15E, 2022ApJ...930L..16E, 2022ApJ...930L..17E}) and the black hole at the heart of the M87 galaxy (\citealt{2019ApJ...875L...1E,2019ApJ...875L...2E,2019ApJ...875L...3E,2019ApJ...875L...4E,2019ApJ...875L...5E,2019ApJ...875L...6E}) have recently been published by the Event Horizon Telescope Collaboration (EHT). These images provide a new probe of the horizon scale physics of supermassive black holes and allow us to study accretion processes and plasma physics in extreme gravitational regimes at an unprecedented resolution. 

Sgr A$^*$ and the black hole in M87 are both low-luminosity active galactic nuclei. They are therefore expected to be hot accretion flows, which are radiatively inefficient (low mass accretion rate) and advection dominated.
Radiatively inefficient accretion flows (RIAFs) comprise a hot accretion plasma in the form of an optically thin and geometrically thick disk (\citealt{1995ApJ...452..710N}, \citealt{2014ARA&A..52..529Y}). Over the past few decades, numerous GRMHD simulation codes have been developed to simulate RIAFs (see e.g., \citealt{Gammie2013}; \citealt{2011A&A...528A.101B,2007ApJS..170..228M,2012MNRAS.426.3241N,2017ComAC...4....1P,2018MNRAS.474L..81L}). In parallel, a variety of general relativistic ray-tracing (GRRT) codes have also been developed to predict the observational appearance of RIAFs from the simulated fluid data (e.g. \citealt{2004A&A...424..733F}; \citealt{2012ApJ...745....1P}; \citealt{2013ApJ...777...13C}; \citealt{2016MNRAS.462..115D}; \citealt{2018MNRAS.475...43M}; \citealt{2022ApJS..262...28W};
\citealt{2025ApJ...985...40S}).

GRMHD and GRRT codes produce libraries of images and movies that can be compared to EHT observations. These comparisons have been used to constrain magnetic field structure, plasma number density and temperature, and the spacetime metric at event horizon scales. EHT observations of Sgr A$^*$ and M87 have been shown to be broadly consistent with theoretical simulations (\citealt{1979A&A....75..228L}; \citealt{2000ApJ...528L..13F}; \citealt{2007ApJS..170..228M}; \citealt{2019ApJ...875L...5E, 2019ApJ...875L...6E}, \citealt{2022ApJ...930L..16E, 2022ApJ...930L..17E}, \citealt{2021ApJ...911L..11E}).

There are, however, discrepancies
between the variability properties of the simulations and the observations \citep{2022ApJ...930L..16E}. 
In particular, EHT analyses of the modulation index of the total light curve (its standard deviation over its mean) have shown that GRMHD simulations often overpredict variability amplitudes.
Interpreting this discrepancy is difficult because the modulation index is a global measure, which is influenced by radiative transfer, emission geometry, gravitational lensing, and Doppler beaming. These additional effects complicate efforts to isolate and calibrate the contribution of the intrinsic fluid simulation variability on the observed light curves. Nevertheless, any physically viable accretion model must reproduce the variability statistics implied by the data. This motivates a more detailed, scale-sensitive analysis, which we aim to provide. 

Turbulence plays an important role in accretion because it enables angular momentum transport from small to large radii. \cite{balbus_powerful_1991} showed that weak magnetic fields can lead to magnetohydrodynamic (MHD) turbulence through the magnetorotational instability (MRI; see also \citealt{1959JETP....9..995V}, \citealt{1960PNAS...46..253C}). Within the context of black hole accretion disks, the strength of the magnetic fields differentiates accretion flows into two regimes. 
A weak magnetic field in a differentially rotating disk results in a SANE flow (standard and normal evolution, \citealt{2003ApJ...592.1042I}). Alternatively, when the magnetic fields are dynamically important, they can oppose the inward ram pressure of the fluid and arrest the accretion flow (magnetically arrested disks, MADs, \citealt{2012MNRAS.426.3241N}). In the present work, we employ various high-resolution, long-duration GRMHD simulations of SANE and MAD (\citealt{2022MNRAS.511.3795N}) flows over a wide range of black hole spins (both prograde and retrograde).

In a turbulent flow, energy cascades across a wide range of spatial scales. MHD turbulence can thus be characterized by a power spectrum that reports the relative importance of eddies of different sizes. Extracting power spectra locally as a function of radius from GRMHD simulations can be a challenge, since spatial variations in the power spectra, the absence of data within the black hole event horizon, limited resolution at specific radii and irregular simulation grids all increase the complexity of the computation. Conceptually, measuring turbulent fluctuations in curved spacetime also requires choices of the observer's frame of reference (see, e.g., \citealt{2025arXiv250904566M}). However, there is a preferred direction in accretion flows, which is determined by the bulk motion of the fluid. We can exploit this preferred direction by assuming that turbulence is ``frozen-in'' the fluid as it moves past the detector, or in our case the simulation grid cells, as stated by Taylor's frozen-in hypothesis (\citealt{1938RSPSA.164..476T}).  By measuring the fluctuations in time as the fluid moves past at different distances from the black hole, and accounting for its velocity, we can approximate the spatial power spectrum from the temporal.

In the present work, we explore the validity of using this approximation for supermassive black hole accretion flows, and employ it to estimate the spatial power spectra of rest-frame number density. We measure temporal power spectra from GRMHD simulations and use Taylor's hypothesis to calculate the corresponding spatial power spectra as a function of radius. We focus on the region close to the black hole since in EHT observations, most of the 1.3~mm emission (the main wavelength the EHT observes) peaks close to the event horizon for both M87 and Sgr A$^*$ (see, e.g., \citealt{2019ApJ...875L...5E, 2025ApJ...985...40S}). While previous work has extracted the power-law indices of turbulence fluctuations in ray-traced images (\citealt{2022ApJ...930L..20G}), we focus on number density fluctuations in the three-dimensional accretion flow.

The paper is structured as follows: In Section \ref{sec:2}, we discuss the applicability of Taylor's hypothesis in MHD turbulence. In Section \ref{sec:3}, we explore the validity of Taylor's hypothesis in black hole accretion flows using GRMHD simulations. In Section \ref{sec:4}, we evaluate spatial power spectra derived from temporal power spectra for one example a non-spinning black hole accretion flow. In Section \ref{sec:5}, we carry out a parameter survey and characterize the power spectra for two magnetic field configurations and various black hole spins. We discuss our results in Section~\ref{sec:6} and conclude in Section \ref{sec:7}.

\section{Taylor's Frozen-In Hypothesis}\label{sec:2}
Under Taylor's frozen-in hypothesis (\citealt{1938RSPSA.164..476T}), temporal fluctuations can be mapped to spatial fluctuations when the turbulence is ``frozen-in'' a faster-moving bulk flow, i.e., when the internal velocity of the fluctuations is negligible compared to the bulk velocity of the flow. Stationary turbulence implies that the total or material derivative of the fluctuations vanishes, i.e.

\begin{equation}
    \frac{{d}\delta \vec{U}}{dt} = \frac{ \partial \delta \vec{U}}{\partial t} + \vec{U} \cdot \nabla \delta \vec{U} = 0,
\end{equation}
and that turbulence is steady in the frame co-moving with the fluid. If $\vec{U} = \vec{U}_{\mathrm{bulk}} + \delta \vec{U}$, one can assume $\delta \vec{U}\cdot\nabla\delta\vec{U}$ to be negligible in the convective term. For an observer in the fluid frame at $\vec{x_s}=\vec{x}+\vec{U}t$, taking space-time Fourier transform of the fluctuations $\delta\vec{U}$ is
\begin{equation}
    \delta \vec{U}(\vec{x},t) = \frac{1}{(2\pi)^4}\int d^3\vec{k}\,d\omega_k\,\delta\tilde{U}(\vec{k},\omega_k)e^{ik\vec{x}-\omega_kt},
    \label{eq:FT}
\end{equation}
and substituting $\vec{x}$ in terms of $\vec{x_s}$ in equation \ref{eq:FT}, we find that 
\begin{equation}    
    \omega_s = \omega_{k} \pm \vec{k}\cdot \vec{U}, 
    \label{eq:TF}
\end{equation}
where $\omega_{k}$ is the internal frequency of turbulence. Assuming that $\omega_k \ll |\vec{k}\cdot\vec{U}|$, equation \ref{eq:TF} maps the original wavenumber $\vec{k}$-spectrum onto an easily detectable stationary observer’s (spacecraft) frequency $\omega_s$-spectrum for fluctuations in velocity.

\begin{figure*}[t]
    \centering
    \includegraphics[width=\textwidth]{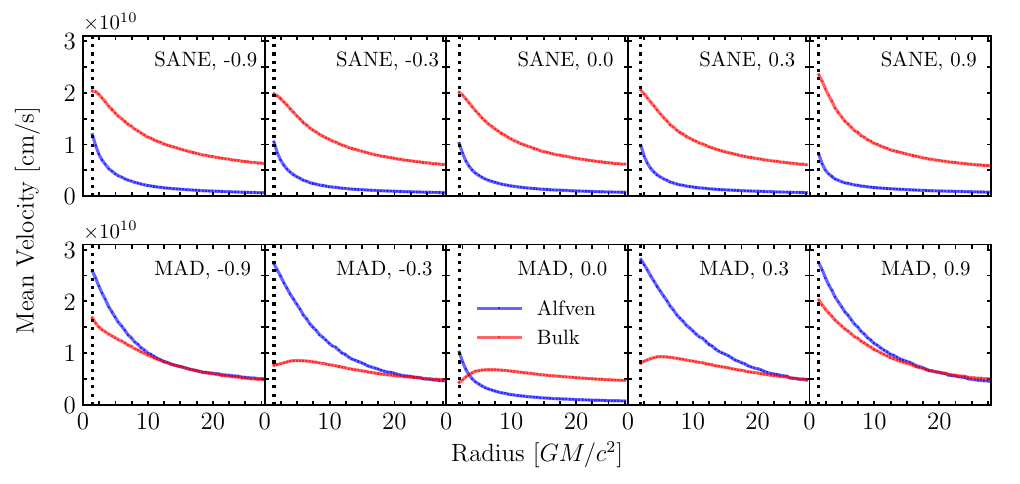}
    \caption{Comparison between Alfv\'en wave velocity and bulk accretion flow velocity in the coordinate normal or stationary frame of reference for SANE \emph{(top row)} and MAD \emph{(bottom row)} flows  with black hole spins ranging from $-0.9$ to $0.9$. Both quantities are averaged azimuthally, within the disk-region defined by polar angles $\theta$ between $45^{\circ}$ and $135^{\circ}$, and within radial bins of width $0.5\ M$. The quantities are also time-averaged over 2000 snapshots, spanning over 20,000 $GM/c^3$. In all SANE flows, the bulk velocity exceeds that of the Alfv\'en wave. The opposite is true in the case of MAD, except for the non-spinning model, for which both velocities are comparable.}
    \label{fig:VC} 
\end{figure*}
 
Taylor's hypothesis has previously been applied in the context of MHD turbulence in solar wind studies using observations from the Parker Solar Probe (\citealt{1995SSRv...73....1T, 2017AGUFMSH21C..02F, 2019EP&S...71...41T, 2021A&A...650A..22P, 2022arXiv220412790V}), which is effectively stationary relative to the higher velocity solar wind.  \citealt{2019EP&S...71...41T} use the Alfv\'en wave speed as the characteristic speed of MHD turbulence; i.e. $\omega_k \approx \vec{{k}} \cdot \vec{U}_{\mathrm{Alf.}}$. When the Alfv\'en wave speed ($\vec{U}_{\mathrm{Alf.}}$) is much lower than the mean bulk flow velocity ($\vec{U}_{\mathrm{bulk}}$; i.e. $\omega_{k} \ll |\vec{k}\cdot\vec{U}_{\mathrm{bulk}}|$ in equation \ref{eq:TF}), one can use the following equation to map the temporal fluctuations to the spatial fluctuations:
\begin{equation}
    \omega_s = \vec{{k}} \cdot \vec{U}_{\mathrm{bulk}}.
    \label{eq:MAP}
\end{equation}
In the above case, the flow is assumed to be ideal MHD, and one must take a closer look at the validity of the hypothesis for compressible and anisotropic flows.

\citealt{2019EP&S...71...41T} show that it is reasonable to use Taylor's hypothesis in MHD turbulence when $\vec{{U}}_{\mathrm{Alf.}} \ll \vec{{U}}_{\mathrm{bulk}}$. This condition would invalidate the hypothesis in regions with higher magnetic field strength, where the two speeds are comparable or the Alfv\'{e}n speed exceeds the bulk velocity. However, \cite{2022arXiv220412790V} develops a more general relation between the spatial and temporal turbulence power spectra that is valid even in regions with high magnetic fields. According to \cite{2022arXiv220412790V}, the following relation can map the frequency spectrum to the wavenumber $k$-spectrum: 
\begin{equation}
    \omega_s = \vec{k} \cdot| \vec{U}_{\mathrm{bulk}} - \vec{U}_{\mathrm{Alf.}} |.
    \label{eq:TFM}
\end{equation}
In the case where $\vec{U}_{\mathrm{bulk}} \gg \vec{U}_{\mathrm{Alf.}}$, this relation reduces to 
equation \ref{eq:MAP}. Otherwise, when $\vec{U}_{\mathrm{bulk}} \ll \vec{U}_{\mathrm{Alf.}}$, we find from equation \ref{eq:TFM} that
\begin{equation}
    \omega_s = \vec{{k}}\cdot\vec{U}_{\mathrm{Alf.}}.
    \label{eq:newMAP}
\end{equation}
Using equations \ref{eq:MAP} and \ref{eq:newMAP} allows us to apply this approximation to most MHD flows with high and low magnetic field strengths. It is important to note that Taylor's hypothesis in the case of MHD turbulence has a number of caveats. It is only applicable for time intervals over which the turbulence can be considered stationary. It also ignores the non-linearity in the convective term which is only reasonable for fast moving flows and small fluctuation amplitudes.

Our aim is to determine how the spatial power spectrum of density fluctuations relates to its temporal power spectrum in different regimes of magnetization, or in other words, with what velocity the density fluctuations pass through a stationary observer. For regions of low magnetic field, the hypothesis is applicable since the velocity fluctuations are carried past the observer by the bulk fluid speed. We extend that approximation to fluctuations in number density ($\delta N$) using the continuity equation in regions of low magnetization. 
For regions of higher magnetic field, the spectrum of $\delta N$ does not follow the spectrum of $\delta \vec{U}$ or $\delta \vec{B}$. In this limit of high magnetization, $\delta N$ is carried by the fast magnetosonic wave, whose speed can be approximated by the Alfv\'en wave speed. 

In the following section, we apply this formulation to our library of GRMHD simulations and explore the applicability of Taylor's hypothesis for fluctuations in number density.

\section{Applicability to black hole accretion flows}\label{sec:3}
 
We consider simulations run using a three-dimensional GRMHD code, {\tt KORAL} (\citealt{2022ApJ...941L..12R, 2022MNRAS.511.3795N, 2014MNRAS.439..503S}), with a resolution of
$288 \times 192 \times 144$ cells in the radial ($r$), polar ($\theta$), and azimuthal ($\phi$) directions, respectively. We use snapshots from the simulations over the time range $10,000 - 30,000\,GM/c^3$, by which point the flow is expected to be in an approximate steady state near the black hole. We include two sets of simulations with different temporal resolutions, one with $\Delta t = 10 \ GM/c^3$, and one with $\Delta t = 0.5 \ GM/c^3$. The simulations evolve the MHD flow through the Kerr metric in spherical Kerr-Schild coordinates. 
We use a simulation grid that is non-uniform and that has been modified to concentrate resolution in both the jet region close to the polar axis and the disk region near the equatorial plane. We set $G=c=1$ throughout this manuscript. 
Our simulation library includes nine different values of black hole spin: 
${a} = -0.9, \, -0.7, \, -0.5, \,-0.3,\,0, \,0.3, \,0.5, $ $ 0.7, \,0.9$, 
where negative spin values indicate that the disk is rotating opposite to the black hole.

The GRMHD simulations evolve the bulk velocity of the flow, magnetic field, and the rest-mass number density at each grid point. The bulk velocity and magnetic field are output in the coordinate frame, while the number density is output in the fluid frame. We compare the mean bulk velocity $\vec{U}_{\mathrm{bulk}}$ and the mean Alfv\'{e}n speed $\vec{U}_{\mathrm{Alf.}}$ to identify which of the mappings between equations \ref{eq:MAP} and \ref{eq:newMAP} to use in the different accretion flows. Moving from MHD to GRMHD introduces a number of complications, such as the necessity of choosing an appropriate reference frame. It is beneficial to choose a reference frame where the temporal and spatial power spectra are easily interpretable. We select the coordinate normal observer frame, with four-velocity
    \begin{equation}
    {n_{\mu}} = \bigg(-\sqrt{\frac{1}{-g^{00}}} , 0 , 0 , 0\bigg),
    \end{equation}

\noindent where $g$ is the Kerr metric.

\par Observers moving with this four-velocity remain fixed relative to the coordinate system, making it a particularly useful frame for analyzing simulations whose output values are at fixed coordinate positions. 
We calculate the classical Alfv\'en speed at each grid point in the simulation using
    \begin{equation}
        U_{\mathrm{Alf.}} = \frac{|B|}{\sqrt{4\pi \rho}},
    \end{equation}
where $|B|$ is the magnitude of the magnetic field as measured in the fluid frame as $B =\sqrt{|b^\mu b_{\mu}|}$, $b^\mu$ is the magnetic induction four-vector for the fluid, and $\rho$ is the mass density in the fluid frame. The permeability of free space, $\mu_0=4\pi$, gives us the classical Alfv\'en speed in centimeter-gram-second (CGS) units.

For regions with very strong magnetic fields, this quantity can exceed the speed of light. Since the Alfv\'en speed may be relativistic in black hole accretion flows, especially MAD flows with strong and ordered magnetic fields, we use the following relativistic correction
    \begin{equation}
        U_{\mathrm{Alf.; rel}} = \frac{U_{\mathrm{Alf.}}/c} {\sqrt{1 + U_{\mathrm{Alf.}} ^ 2/c^2}},
    \end{equation}
where $c$ is the speed to light in CGS units, which ensures $U_{\mathrm{Alf.; rel}}<c$.

\begin{figure*}[t]
    \centering
    \includegraphics[width=\textwidth]{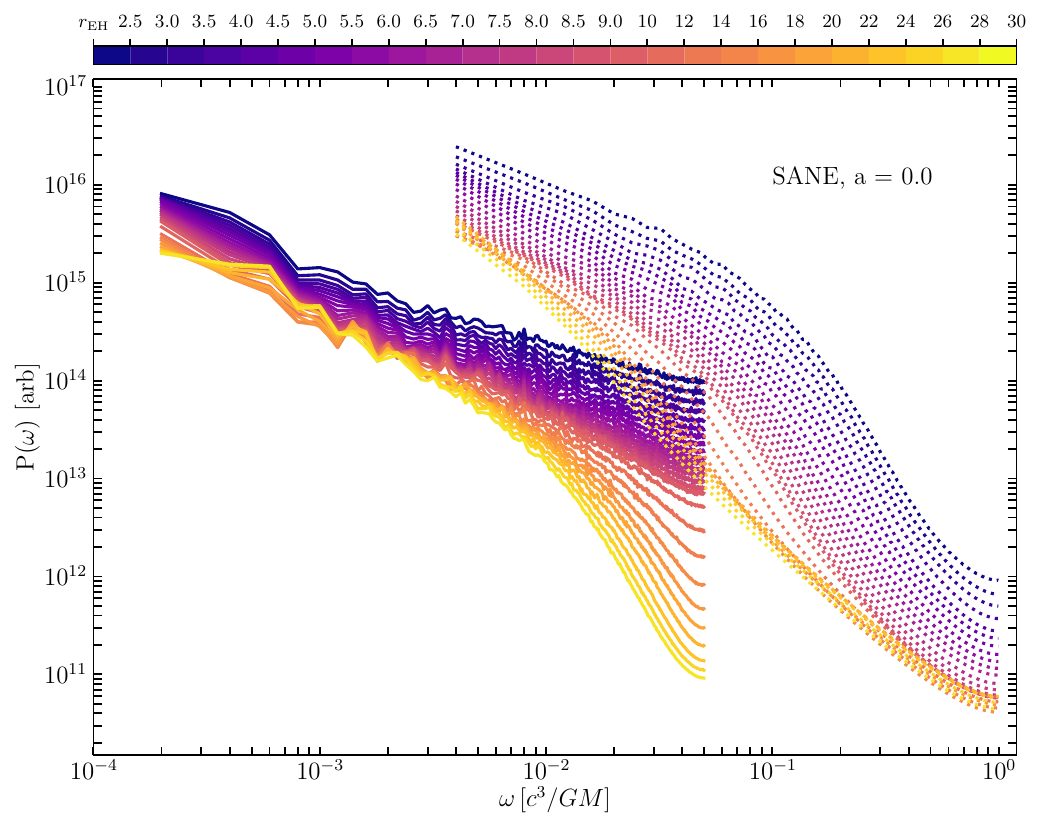}
    \caption{Azimuthally- and disk-averaged temporal power spectra of number density fluctuations for a non-spinning SANE flow. We show simulations with low cadence ($\Delta T = 10 \ GM/c^3$, \emph{solid}) and high cadence ($\Delta T = 0.5 \ GM/c^3$, \emph{dotted}). Each curve corresponds to the average one-dimensional temporal power spectrum ($\omega$-spectrum) of number density fluctuations within a range of radii from the black hole. Colors denote the radial ranges in units of $GM/c^2$.}
    \label{fig:temp_comp_SANEa9} 
\end{figure*}

\begin{figure*}[t]
    \centering
    \includegraphics[width=\textwidth]{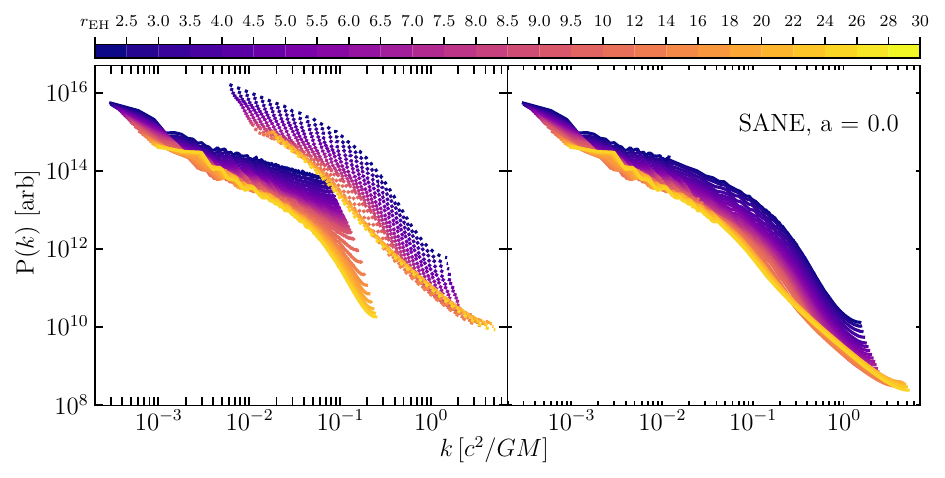}
    \caption{Azimuthally- and disk-averaged spatial power spectra of number density fluctuations for a non-spinning SANE flow. (\emph{left}) Each temporal power spectrum from Figure \ref{fig:temp_comp_SANEa9} is mapped to its corresponding spatial spectrum  following equation \ref{eq:MAP}. (\emph{right}) Normalized spectra with lower frequency data points of the higher resolution $0.5 \ GM/c^3$ simulations prioritized, i.e., for frequencies where the two simulations overlap, we keep data points from the higher-cadence simulation.}
    \label{fig:spat_comp_SANEa9} 
\end{figure*}

In Figure \ref{fig:VC}, we compare the mean bulk speed to the Alfv\'en speed for a representative subset of simulations. We average these velocities azimuthally in $\phi$, in $\theta$ within a fixed angle of $45^{\circ}$ to $135^{\circ}$ (which we define as the disk region), and in time over the full duration of each simulation ($20,\!000 \ GM/c^3$ and $2000 \ GM/c^3$ for the low and high cadence simulations, respectively). 

In SANE flows, $|\vec{U}_{\mathrm{bulk}}| > |\vec{U}_{\mathrm{Alf.}}|$, allowing us to employ equation \ref{eq:MAP} to convert temporal power spectra into spatial power spectra. In this case, the flows lie in the limit of low magnetization, and the density fluctuations are carried by the mean bulk velocity. In spinning MAD flows, $|\vec{U}_{\mathrm{Alf.}}| > |\vec{U}_{\mathrm{bulk}}|$, motivating the use of equation \ref{eq:newMAP} to extract the $k$-spectrum. Since MAD flows lie in the limit of high magnetization, the density fluctuations are carried by the Alfv\'en speed. For the non-spinning MAD flow, the two velocities are comparable and we cannot use either mapping. Hence, we do not consider this case for the analysis. We apply this methodology to the rest-frame number density power spectrum in the following section.

\section{Converting Temporal to Spatial power spectra}\label{sec:4}

The properties of turbulence are often quantified in terms of the power spectral density, which describes the distribution of energy at different spatial scales.
The theory of turbulent power spectra has been worked out for several different physical scenarios. In the simplest case of isotropic and homogeneous turbulence, \cite{1941DoSSR..32...16K} showed that the distribution of kinetic energy across wavenumber $k$ is
\begin{equation}
    E_u(k) = K\epsilon_u^{2/3}k^{-5/3},
\end{equation}
where $K$ is the Kolmogorov constant and $\epsilon_u$ is the dissipation rate of kinetic energy.

Extensions to the Kolmogorov model have been proposed, especially in the context of magnetized plasmas where the magnetic field can introduce a preferred direction. When the strength of the mean magnetic field is sufficiently high, the turbulence is effectively two-dimensional and typically yields 
a $k^{-3/2}$ spectrum due to uncorrelated collisions of Alfv\'{e}n wavepackets \citep{1963AZh....40..742I,1965PhFl....8.1385K}. Three-dimensional magnetized turbulence is also possible, although non-linear couplings and resonant interactions often cause it to become quasi-two-dimensional (\citealt{1983JPlPh..29..525S}; see also \citealt{1995ApJ...438..763G} for anisotropic power spectra in the context of symmetric Alfv\'{e}n waves). Note that the power spectrum is often determined by more than just the properties of the turbulent eddy structure. In the case of an accretion flow, for example, disk modes (perhaps due to instabilities or resonances), large scale trends with radius, and shocks can all influence the power spectrum. 

We characterize the power spectrum of rest-frame number density as seen in time-dependent black hole accretion simulations. In particular, we measure the power spectrum as a function of distance from the black hole, black hole spin, and the strength of the event-horizon-scale magnetic field (SANE vs.~MAD). The power spectra we report are computed and averaged over the entire disk region as defined in Section \ref{sec:3} in order to estimate properties of the global variability. We compute the power spectrum of rest-frame number density since number density is a scalar and is therefore easier to work with compared to directional quantities like velocity or magnetic field.\footnote{In GRMHD simulations, frame transformations of directional quantities such as the magnetic field and velocity are more involved and non-trivial, while the rest frame number density, being an intrinsic scalar quantity is invariant under Lorentz boosts and coordinate transformations.}

Since we are motivated by the recent EHT observations at 1.3~mm, we focus on regions close to the black hole ($r<30 \ M$), where the majority of the observed emission is produced. We record the time series of number density at each grid point in the simulation and Fourier transform it to obtain the associated temporal power spectrum. We then average the power spectra in $\theta$ and $\phi$ as described above, as well as over radial bins of width $0.5\ M$ for $r_{\rm eh} < r < 10\ M$ and bins of width $2\ M$ for $r > 10\ M$. Figure~\ref{fig:temp_comp_SANEa9} shows an example of the temporal power spectra for a non-spinning SANE flow using data from both the low-cadence and high-cadence simulations.

We apply Taylor's frozen hypothesis to the temporal power spectra in Figure~\ref{fig:temp_comp_SANEa9} using equation~\ref{eq:MAP}, since in SANE simulations the flow velocity is well above the Alfv\'en velocity (see Figure~\ref{fig:VC}). Because the flow velocity changes with radius, we independently compute $|\vec{U}_{\rm bulk} |$ for each radial bin by averaging over $\theta$ and $\phi$ within the bin. The spatial power spectrum can then be computed as

\begin{equation}
    P(k) = P(\omega)|\vec{U}_{\rm bulk}|, 
\end{equation}
where $k = \omega / |\vec{U}_{\rm bulk} |$.

We show the resulting spatial power spectra in Figure~\ref{fig:spat_comp_SANEa9}.

Equation~\ref{eq:MAP} can be used for all SANE models, since the bulk flow velocity always exceeds the Alfv\'en speed for these simulations (see Figure~\ref{fig:VC}). In contrast, for spinning MAD models, $|\vec{U}_{\mathrm{bulk}}| > |\vec{U}_{\mathrm{Alf.}}|$ and we use equation~\ref{eq:newMAP} to map the temporal spectrum to the wavenumber spectrum, so that
\begin{equation}
    P(k) = P(\omega)|\vec{U}_{\mathrm{Alf.}}|.
\end{equation}

\begin{figure*}[t]
    \centering
    \includegraphics[width=1\textwidth]{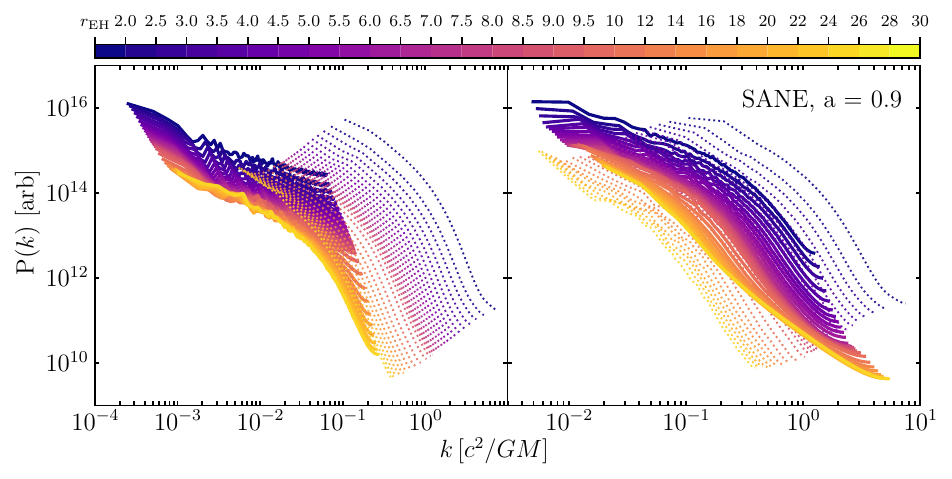}
    \caption{Azimuthal ($\phi$) power spectra of number density fluctuations averaged over time and within the disk region for a SANE simulation with black hole spin $a = 0.9$ \emph{(dotted)}. We compare the $\phi$ spectra to those from Taylor’s hypothesis \emph{(solid)} for the low-cadence ($10\ GM/c^3$, \emph{left}) and high cadence ($0.5\ GM/c^3$, \emph{right}) SANE simulation.}
    \label{fig:phi_SANEa9} 
\end{figure*}

The range of temporal frequencies that can be probed in a simulation depends only on the cadence and duration of the simulation and is therefore constant across the domain. However, since the mapping between the temporal and spatial power spectra depends on velocity (see equations~\ref{eq:MAP} and \ref{eq:newMAP}), the range of spatial frequencies that can be probed varies with radius. In Figure~\ref{fig:spat_comp_SANEa9}, we show the spatial power spectra computed from both the high and low cadence simulations across all radii. We rescale the amplitude of the power spectra so that they match and prioritize data points from the high cadence simulation to lessen the influence of aliasing. These power spectra show that there is more coherent power at larger spatial scales (or lower frequencies) and that the power drops off slowly at low frequencies and then more rapidly after a break frequency. The break frequency tends to decrease with distance from the black hole.

As a qualitative check of Taylor's frozen-in hypothesis, we compare the power spectra in $\phi$ to the result from the hypothesis in Figure \ref{fig:phi_SANEa9}. Although the power spectra in $\phi$ measures the spatial number density profiles more directly, it is limited by the spatial resolution of the simulation and ignores radial motion. We perform this comparison for a SANE flow where radial velocities are expected to be small. We find that the spectra in $\phi$ are broadly consistent with the spectra obtained with Taylor's hypothesis. The slopes and dynamical ranges of the power spectra in $\phi$ match more closely with the power spectra of the higher-cadence simulations.

\section{Parameter Survey}\label{sec:5}

\begin{figure*}[t]
    \centering
    \includegraphics[width=\textwidth]{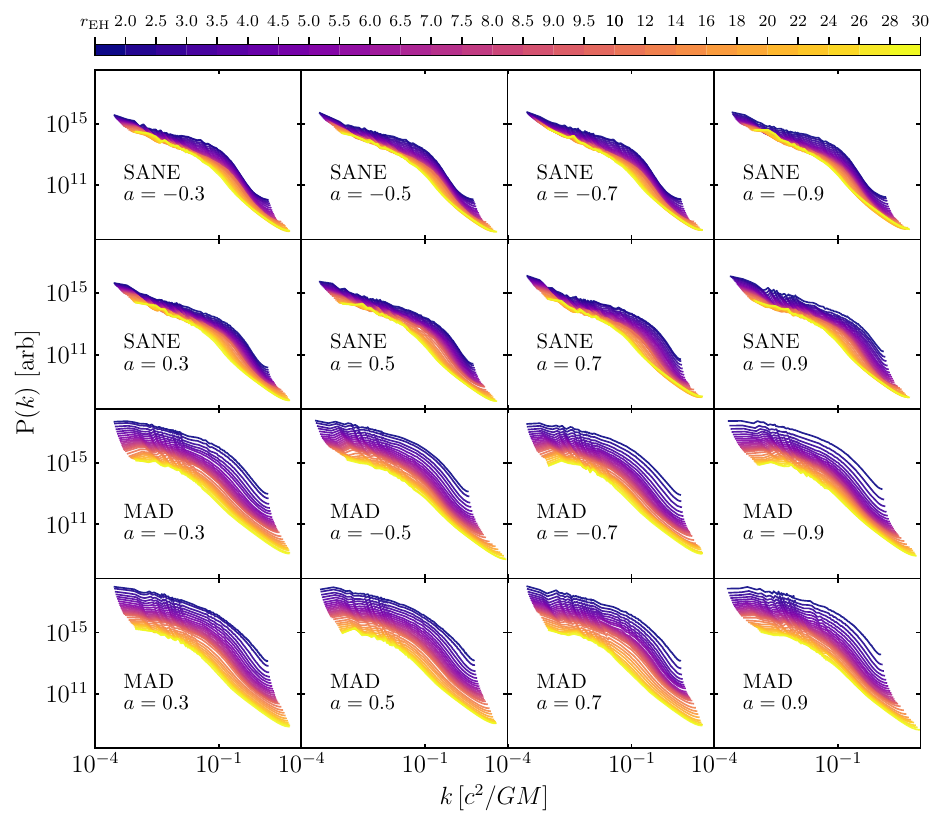}
    \caption{Azimuthally- and disk-averaged spatial power spectra of number density fluctuations from MAD and SANE simulations for different black hole spins. The spectra are normalized and we prioritize lower frequency data points of the higher resolution $0.5 \ GM/c^3$ simulations as shown in Figure \ref{fig:spat_comp_SANEa9}.}
    \label{fig:PS} 
\end{figure*}

\begin{figure}[t]
    \centering
    \includegraphics[width=\columnwidth]{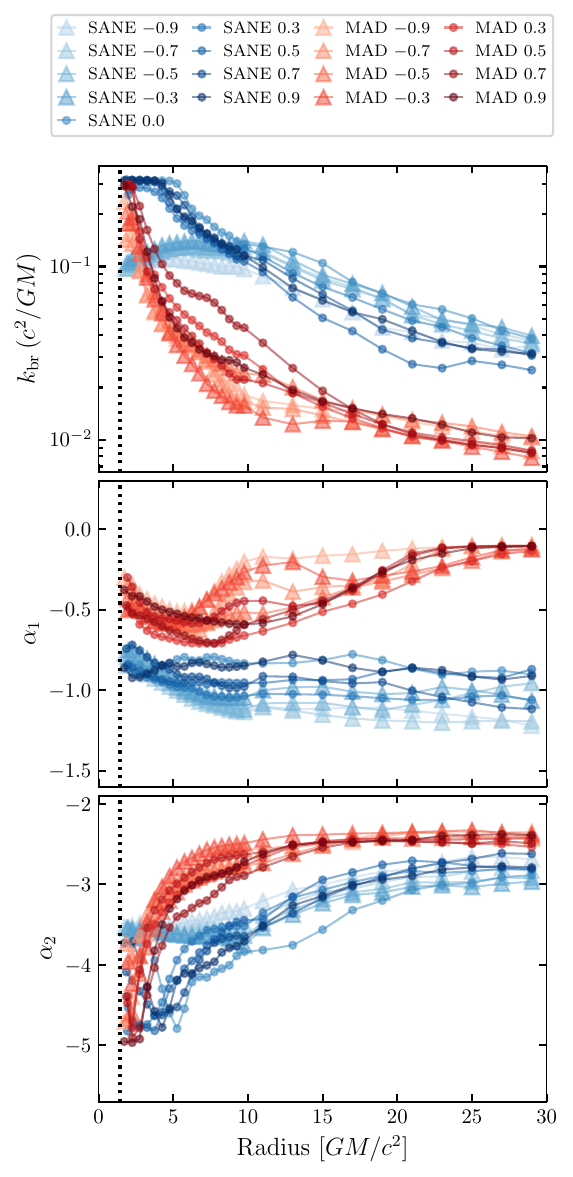}
    \caption{Results from fitting the power spectra profiles to a Beuermann profile (equation \ref{BPL}) with break frequency (\emph{top}), slope before the break frequency $\alpha_1$ (\emph{middle}), and slope after the break frequency $\alpha_2$ (\emph{bottom}).}
    \label{fig:fits} 
\end{figure}

We apply the analysis from Section \ref{sec:4} to SANE and MAD flows with spins $-0.9$ to $0.9$. We map between the measured temporal power spectra and the spatial ones using the same procedure used for Figure \ref{fig:spat_comp_SANEa9}. Following Figure \ref{fig:VC}, we use the bulk flow velocity for SANE flows and the Alfv\'en speed for MAD flows. The results of the parameter survey are shown in Figure \ref{fig:PS}. We characterize the power spectra with a broken power law (Beuermann profile; \citealp{1999A&A...352L..26B})
\begin{equation}\label{BPL}
    P = f \; \Biggl(\biggl(\frac{k}{k_{\mathrm{br}}}\biggr)^{-h\alpha_1} + \biggl(\frac{k}{k_{\mathrm{br}}}\biggr)^{-h\alpha_2}\Biggr)^{-1/h},
\end{equation}
where the multiplicative scaling factor $f$ sets the power density scale, $k_{\mathrm{br}}$ is the break frequency, $h$ is the sharpness of the transition between the two power laws, and $\alpha_1$ and $\alpha_2$ are the two slopes. 

We fit the spatial power spectra to equation~\ref{BPL} using {\tt emcee}, a python implementation of a Markov chain Monte Carlo algorithm (\citealt{emcee}). To avoid aliasing artifacts in the power spectrum, we implement a radially dependent high-frequency cutoff, with the cutoff frequency increasing proportional to the radius from the black hole. We show the results of these fits in Figure~\ref{fig:fits}. 

We find that the break frequency, $k_{\mathrm{br}}$, ranges from $\approx 0.01$ to $\approx 0.3$ and tends to decrease with distance from the black hole, but MAD and SANE show distinctly different trends. In MAD flows, the break frequency falls off rapidly until $\approx 5 - 10 \ M$ and then tapers off, while in SANE flows $k_{\mathrm{br}}$ descends less rapidly. In SANE flows, we find that the retrograde spins separate from the prograde spins and, within $10\ M$, the break frequency is larger for non-spinning and prograde flows.

Physically, the break frequency separates the power spectrum into two slopes, which probe two different regimes of turbulent dynamics in the accretion flow. Given two fixed slopes, a higher break frequency implies a relatively lower power at larger frequencies compared to smaller frequencies and vice versa. Higher break frequencies close to the black hole imply more small-scale density fluctuations. With increasing distance from the black hole, a decreasing break frequency indicates an increase in relative large-scale fluctuations, which could be due to relativistic effects that become weaker with larger radii.

The slope of the spectra at lower frequencies, $\alpha_1$, is shallow, ranging from $\approx - 0.1$ to $\approx - 1.25$. We find that $\alpha_1$ increases slightly with radius for MAD flows and is roughly constant for SANE flows. The slope at higher frequencies, $\alpha_2$, is steeper than $\alpha_1$, ranging from $\approx -2.4$ to $\approx -5$ and saturating at $\approx -2.5$ for MAD and $\approx -3$ for SANE. For MAD flows, $\alpha_2$ increases steeply with radius till $10 \ M$ and then continues to increase more gradually. For SANE flows, in the prograde models, $\alpha_2$ shows a dip until $\approx 5 \ M$ and then increases gradually, whereas retrograde flows exhibit a more consistent gradual increase in $\alpha_2$ with radius. Both $\alpha_1$ and $\alpha_2$ appear to display systematic dependencies on both the magnitude and sign of the black hole spin, but these trends are more difficult to interpret.

\section{Discussion}\label{sec:6}

An important discrepancy between EHT observations and GRMHD models is the difference in the overall amplitude of the variability computed via the modulation index of the total light curve. Connecting the spatially and temporally resolved variability measured in this paper to this global observational statistic is nontrivial, due to the influence of radiative transfer, gravitational lensing, and relativistic Doppler effects, which differentially weight regions of the flow. Nonetheless, any physically realistic model of the accretion flow should reproduce the intrinsic variability statistics we have characterized.

\begin{figure*}[t]
    \centering
    \includegraphics[width=\textwidth]{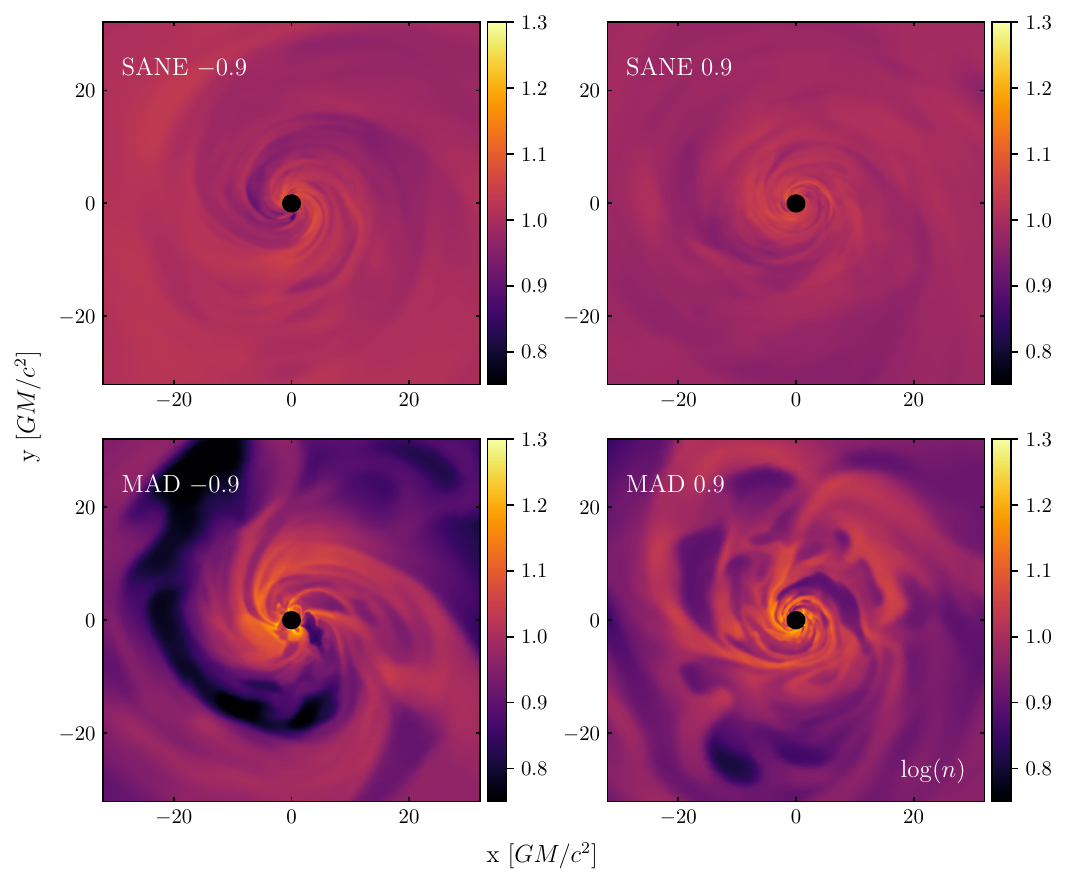}
    \caption{Snapshots of GRMHD flows showing rest mass number density (logarithmic) in the equatorial plane of the accretion disk. SANE flows for black hole spins $-0.9$ and $0.9$ (\textit{top row}) show a distribution with organized accretion strands, while MAD flows for black hole spins $-0.9$ and $0.9$ (\textit{bottom row}) have large voids and non-uniform accretion strands.}
    \label{fig:eq} 
\end{figure*}

Our findings are broadly consistent with qualitative trends observed in time-dependent GRMHD simulations of black hole accretion flows. In particular, as illustrated in Figure \ref{fig:eq}, both MAD flows and retrograde SANE flows frequently exhibit enhanced spatial coherence at large scales, which manifests as persistent, organized accretion strands. Such structures contribute excess power at low spatial frequencies (small $k$), increasing the ratio of large- to small-scale power in the inner regions of the accretion flow where these features are most pronounced.

Our results are consistent with the trends seen in GRMHD simulations, i.e., that retrograde SANE flows and MAD models, which preferentially exhibit stronger low-$k$ power, tend to yield higher modulation indices in image-based analyses \citep{2022ApJ...930L..16E}. We defer a detailed exploration of the connection between intrinsic variability and observable signatures to future work.

Although MAD and SANE flows have different variability characteristics, the trends do not obviously correlate in any simple way with the steady-state scalar quantities in the GRMHD.
A preliminary geometric comparison reveals no clear, consistent association between the break frequency and features such as the disk-corona boundary, magnetic field gradients, or plasma properties like magnetization or plasma-$\beta$, the ratio between gas and magnetic pressure. In fact, in high-spin MADs, the disk is geometrically thinner than in SANE flows, yet the associated variability length scale ($1/k_{\rm br}$) is larger. Similarly, the disk scale height, estimated from the full width at half maximum of the density profile, shows little variation between configurations and does not follow the observed spectral differences.

These results suggest that the break frequency and spectral slopes may be shaped by more global or dynamical processes 
that are not easily captured by instantaneous scalar quantities. For instance, the transition from large- to small-scale power may reflect the coherence length of turbulence in multiple dimensions or a transition in the dominant mode of energy transport. It is also possible that the break reflects the spatial coherence of the accretion flow rather than a direct physical scale. For example, in both MAD and retrograde SANE flows, the presence of persistent, large-scale accretion streams and coherent structures may drive enhanced low-$k$ power. These results are difficult to reconcile with the turbulent theory described in Section~\ref{sec:4}; if anything, the enhanced coherence and large-scale power seen in MADs might suggest a lower-dimensional cascade, even though the statistical properties of the flow do not exactly align with either the 2D or 3D predictions. This suggests that accretion-driven turbulence in strongly relativistic and magnetized environments may operate in a regime that is not well described by classical MHD turbulence models. We leave a more detailed investigation to future work.

\section{Summary}\label{sec:7}

We present
a framework to characterize the fluctuations in number density in black hole accretion flows at different distances from the black hole. 
We use Taylor's frozen-in hypothesis to compute the spatial power spectra of number density from their temporal power spectra.The spatial power spectra allow us to quantify how number density fluctuates on different length scales and can be a useful tool to understand variability in accretion flows like the ones observed by the EHT.

Taylor’s hypothesis leverages the assumption that the spatial variations of interest change slowly and are therefore effectively ``frozen’’ into a faster-moving flow. By subtracting out the bulk fluid motion, it is then possible to recover the structure of the spatial variations. In order to perform this procedure, it is necessary to first identify the characteristic speed of the faster-moving flow. In our suite of simulations, we find that the fluid velocity carries the fluctuations in SANE flows and argue that the Alfv\'en speed is typically a good proxy for the rate at which they are carried in MAD flows. In the non-spinning MAD flow, the two speeds are comparable and we exclude that simulation from our analysis.

To capture a broader range of dynamical timescales, we include simulations at two temporal resolutions: a high-resolution cadence of $0.5 \: GM/c^3$ spanning $1000 \ GM/c^2$ and a low-resolution cadence of $10 \: GM/c^3$ spanning $20,\!000 \ GM/c^2$. For each simulation, we compute azimuthally and disk-averaged power spectra and velocity profiles at different radii to measure the time-averaged variability of number density fluctuations. We obtain the spatial power spectra using Taylor's hypothesis. Our results show that number density fluctuations follow a broken power-law with distinct slopes across different spatial and temporal scales separated by a break frequency. We fit a broken power law to the spectra and explore how the break frequencies and slopes change as a function of radial distance from the black hole.

Across all simulations, we find that the break frequency decreases with distance from the black hole. The rate at which the break frequency decreases is higher in MAD flows. In SANE flows, we also note a dependence on the sign of the black hole spin, with prograde spins showing a more rapid decrease in break frequency with radius as compared to retrograde spins.

The spectral slopes $\alpha_1$ and $\alpha_2$ also have some notable trends. The low-frequency slope, $\alpha_1$, is characteristically steeper for SANE flows than MAD flows. The $\alpha_2$ for SANE prograde models dips close to the black hole before increasing gradually. Retrograde SANE models increase gradually with distance. All SANE models saturate at $\approx -3$. MAD models increase and saturate at $\approx -2.5$. 

Our results provide a framework for characterizing variability in these accretion flows via the
break frequency and spectral slopes of the power spectrum, which can inform interpretations of interferometric data from the EHT. 
Since our analysis focuses on the intrinsic flow rather than its radiative signatures, the spectral break frequencies and slopes can be interpreted as inherent characteristics of the accretion dynamics. 
Future work is necessary to connect more directly to observables and explore the role of additional physics such as radiative transfer and electron thermodynamics.

\vspace{0.2in}
The authors thank M. Faraco de Medeiros and J.~Stone for fruitful discussions. The authors thank R.~Narayan, A.~Ricarte, and A.~Chael for providing the suite of KORAL GRMHD simulations analyzed in this paper.  
P.\;H.\ gratefully acknowledges support from DST-INSPIRE for the INSPIRE scholarship no. MS18198 and thanks the Institute for Advanced Study for their hospitality during the course of this work. P.\;H.\ also thanks University of Colorado Boulder, where a part of this work was conducted.
L.\;M.\ gratefully acknowledges support from a NASA Hubble Fellowship Program, Einstein Fellowship under award number HST-HF2-51539.001-A, an NSF Astronomy and Astrophysics 
Postdoctoral Fellowship under award no. AST-1903847, and NSF AST-2407810. 
G.N.W.~was supported by the Taplin Fellowship and the Princeton Gravity Initiative.

\bibliography{main}

\begin{thebibliography}{}
\expandafter\ifx\csname natexlab\endcsname\relax\def\natexlab#1{#1}\fi

\bibitem[{Balbus \& Hawley(1991)}]{balbus_powerful_1991}
Balbus, S.~A., \& Hawley, J.~F. 1991, The Astrophysical Journal, 376, 214

\bibitem[{{Beuermann} {et~al.}(1999){Beuermann}, {Hessman}, {Reinsch}, {Nicklas}, {Vreeswijk}, {Galama}, {Rol}, {van Paradijs}, {Kouveliotou}, {Frontera}, {Masetti}, {Palazzi}, \& {Pian}}]{1999A&A...352L..26B}
{Beuermann}, K., {Hessman}, F.~V., {Reinsch}, K., {et~al.} 1999, \aap, 352, L26

\bibitem[{{Bucciantini} \& {Del Zanna}(2011)}]{2011A&A...528A.101B}
{Bucciantini}, N., \& {Del Zanna}, L. 2011, \aap, 528, A101

\bibitem[{{Chan} {et~al.}(2013){Chan}, {Psaltis}, \& {{\"O}zel}}]{2013ApJ...777...13C}
{Chan}, C.-k., {Psaltis}, D., \& {{\"O}zel}, F. 2013, \apj, 777, 13

\bibitem[{{Chandrasekhar}(1960)}]{1960PNAS...46..253C}
{Chandrasekhar}, S. 1960, Proceedings of the National Academy of Science, 46, 253

\bibitem[{{Dexter}(2016)}]{2016MNRAS.462..115D}
{Dexter}, J. 2016, \mnras, 462, 115

\bibitem[{{EHT MWL Science Working Group} {et~al.}(2021){EHT MWL Science Working Group}, {Algaba}, {Anczarski}, {Asada}, {Balokovi{\'c}}, {Chandra}, {Cui}, {Falcone}, {Giroletti}, {Goddi}, {Hada}, {Haggard}, {Jorstad}, {Kaur}, {Kawashima}, {Keating}, {Kim}, {Kino}, {Komossa}, {Kravchenko}, {Krichbaum}, {Lee}, {Lu}, {Lucchini}, {Markoff}, {Neilsen}, {Nowak}, {Park}, {Principe}, {Ramakrishnan}, {Reynolds}, {Sasada}, {Savchenko}, {Williamson}, {Event Horizon Telescope Collaboration}, {Akiyama}, {Alberdi}, {Alef}, {Anantua}, {Azulay}, {Baczko}, {Ball}, {Barrett}, {Bintley}, {Benson}, {Blackburn}, {Blundell}, {Boland}, {Bouman}, {Bower}, {Boyce}, {Bremer}, {Brinkerink}, {Brissenden}, {Britzen}, {Broderick}, {Broguiere}, {Bronzwaer}, {Byun}, {Carlstrom}, {Chael}, {Chan}, {Chatterjee}, {Chatterjee}, {Chen}, {Chen}, {Chesler}, {Cho}, {Christian}, {Conway}, {Cordes}, {Crawford}, {Crew}, {Cruz-Osorio}, {Davelaar}, {de Laurentis}, {Deane}, {Dempsey}, {Desvignes}, {Dexter}, {Doeleman}, {Eatough}, {Falcke}, {Farah},
  {Fish}, {Fomalont}, {Ford}, {Fraga-Encinas}, {Friberg}, {Fromm}, {Fuentes}, {Galison}, {Gammie}, {Garc{\'\i}a}, {Gentaz}, {Georgiev}, {Gold}, {G{\'o}mez}, {G{\'o}mez-Ruiz}, {Gu}, {Gurwell}, {Hecht}, {Hesper}, {Ho}, {Ho}, {Honma}, {Huang}, {Huang}, {Hughes}, {Ikeda}, {Inoue}, {Issaoun}, {James}, {Jannuzi}, {Janssen}, {Jeter}, {Jiang}, {Jim{\'e}nez-Rosales}, {Johnson}, {Jung}, {Karami}, {Karuppusamy}, {Kettenis}, {Kim}, {Kim}, {Kim}, {Koay}, {Kofuji}, {Koch}, {Koyama}, {Kramer}, {Kramer}, {Kuo}, {Lauer}, {Levis}, {Li}, {Li}, {Lindqvist}, {Lico}, {Lindahl}, {Liu}, {Liu}, {Liuzzo}, {Lo}, {Lobanov}, {Loinard}, {Lonsdale}, {MacDonald}, {Mao}, {Marchili}, {Marrone}, {Marscher}, {Mart{\'\i}-Vidal}, {Matsushita}, {Matthews}, {Medeiros}, {Menten}, {Mizuno}, {Mizuno}, {Moran}, {Moriyama}, {Moscibrodzka}, {M{\"u}ller}, {Musoke}, {Mej{\'\i}as}, {Nagai}, {Nagar}, {Nakamura}, {Narayan}, {Narayanan}, {Natarajan}, {Nathanail}, {Neri}, {Ni}, {Noutsos}, {Okino}, {Olivares}, {Ortiz-Le{\'o}n}, {Oyama}, {{\"O}zel}, {Palumbo},
  {Patel}, {Pen}, {Pesce}, {Pi{\'e}tu}, {Plambeck}, {Popstefanija}, {Porth}, {P{\"o}tzl}, {Prather}, {Preciado-L{\'o}pez}, {Psaltis}, {Pu}, {Rao}, {Rawlings}, {Raymond}, {Rezzolla}, {Ricarte}, {Ripperda}, \& {Roelofs}}]{2021ApJ...911L..11E}
{EHT MWL Science Working Group}, {Algaba}, J.~C., {Anczarski}, J., {et~al.} 2021, \apjl, 911, L11

\bibitem[{{Event Horizon Telescope Collaboration} {et~al.}(2019{\natexlab{a}}){Event Horizon Telescope Collaboration}, {Akiyama}, {Alberdi}, {Alef}, {Asada}, {Azulay}, {Baczko}, {Ball}, {Balokovi{\'c}}, {Barrett}, \& et~al.}]{2019ApJ...875L...1E}
{Event Horizon Telescope Collaboration}, {Akiyama}, K., {Alberdi}, A., {et~al.} 2019{\natexlab{a}}, \apjl, 875, L1

\bibitem[{{Event Horizon Telescope Collaboration} {et~al.}(2019{\natexlab{b}}){Event Horizon Telescope Collaboration}, {Akiyama}, {Alberdi}, {Alef}, {Asada}, {Azulay}, {Baczko}, {Ball}, {Balokovi{\'c}}, {Barrett}, \& et~al.}]{2019ApJ...875L...2E}
---. 2019{\natexlab{b}}, \apjl, 875, L2

\bibitem[{{Event Horizon Telescope Collaboration} {et~al.}(2019{\natexlab{c}}){Event Horizon Telescope Collaboration}, {Akiyama}, {Alberdi}, {Alef}, {Asada}, {Azulay}, {Baczko}, {Ball}, {Balokovi{\'c}}, {Barrett}, \& et~al.}]{2019ApJ...875L...3E}
---. 2019{\natexlab{c}}, \apjl, 875, L3

\bibitem[{{Event Horizon Telescope Collaboration} {et~al.}(2019{\natexlab{d}}){Event Horizon Telescope Collaboration}, {Akiyama}, {Alberdi}, {Alef}, {Asada}, {Azulay}, {Baczko}, {Ball}, {Balokovi{\'c}}, {Barrett}, \& et~al.}]{2019ApJ...875L...4E}
---. 2019{\natexlab{d}}, \apjl, 875, L4

\bibitem[{{Event Horizon Telescope Collaboration} {et~al.}(2019{\natexlab{e}}){Event Horizon Telescope Collaboration}, {Akiyama}, {Alberdi}, {Alef}, {Asada}, {Azulay}, {Baczko}, {Ball}, {Balokovi{\'c}}, {Barrett}, \& et~al.}]{2019ApJ...875L...5E}
---. 2019{\natexlab{e}}, \apjl, 875, L5

\bibitem[{{Event Horizon Telescope Collaboration} {et~al.}(2019{\natexlab{f}}){Event Horizon Telescope Collaboration}, {Akiyama}, {Alberdi}, {Alef}, {Asada}, {Azulay}, {Baczko}, {Ball}, {Balokovi{\'c}}, {Barrett}, \& et~al.}]{2019ApJ...875L...6E}
---. 2019{\natexlab{f}}, \apjl, 875, L6

\bibitem[{{Event Horizon Telescope Collaboration} {et~al.}(2022{\natexlab{a}}){Event Horizon Telescope Collaboration}, {Akiyama}, {Alberdi}, {Alef}, {Algaba}, {Anantua}, {Asada}, {Azulay}, {Bach}, {Baczko}, {Ball}, {Balokovi{\'c}}, {Barrett}, {Baub{\"o}ck}, {Benson}, {Bintley}, {Blackburn}, {Blundell}, {Bouman}, {Bower}, {Boyce}, {Bremer}, {Brinkerink}, {Brissenden}, {Britzen}, {Broderick}, {Broguiere}, {Bronzwaer}, {Bustamante}, {Byun}, {Carlstrom}, {Ceccobello}, {Chael}, {Chan}, {Chatterjee}, {Chatterjee}, {Chen}, {Chen}, {Cheng}, {Cho}, {Christian}, {Conroy}, {Conway}, {Cordes}, {Crawford}, {Crew}, {Cruz-Osorio}, {Cui}, {Davelaar}, {De Laurentis}, {Deane}, {Dempsey}, {Desvignes}, {Dexter}, {Dhruv}, {Doeleman}, {Dougal}, {Dzib}, {Eatough}, {Emami}, {Falcke}, {Farah}, {Fish}, {Fomalont}, {Ford}, {Fraga-Encinas}, {Freeman}, {Friberg}, {Fromm}, {Fuentes}, {Galison}, {Gammie}, {Garc{\'\i}a}, {Gentaz}, {Georgiev}, {Goddi}, {Gold}, {G{\'o}mez-Ruiz}, {G{\'o}mez}, {Gu}, {Gurwell}, {Hada}, {Haggard}, {Haworth},
  {Hecht}, {Hesper}, {Heumann}, {Ho}, {Ho}, {Honma}, {Huang}, {Huang}, {Hughes}, {Ikeda}, {Impellizzeri}, {Inoue}, {Issaoun}, {James}, {Jannuzi}, {Janssen}, {Jeter}, {Jiang}, {Jim{\'e}nez-Rosales}, {Johnson}, {Jorstad}, {Joshi}, {Jung}, {Karami}, {Karuppusamy}, {Kawashima}, {Keating}, {Kettenis}, {Kim}, {Kim}, {Kim}, {Kim}, {Kino}, {Koay}, {Kocherlakota}, {Kofuji}, {Koch}, {Koyama}, {Kramer}, {Kramer}, {Krichbaum}, {Kuo}, {La Bella}, {Lauer}, {Lee}, {Lee}, {Leung}, {Levis}, {Li}, {Lico}, {Lindahl}, {Lindqvist}, {Lisakov}, {Liu}, {Liu}, {Liuzzo}, {Lo}, {Lobanov}, {Loinard}, {Lonsdale}, {Lu}, {Mao}, {Marchili}, {Markoff}, {Marrone}, {Marscher}, {Mart{\'\i}-Vidal}, {Matsushita}, {Matthews}, {Medeiros}, {Menten}, {Michalik}, {Mizuno}, {Mizuno}, {Moran}, {Moriyama}, {Moscibrodzka}, {M{\"u}ller}, {Mus}, {Musoke}, {Myserlis}, {Nadolski}, {Nagai}, {Nagar}, {Nakamura}, {Narayan}, {Narayanan}, {Natarajan}, {Nathanail}, {Fuentes}, {Neilsen}, {Neri}, {Ni}, {Noutsos}, {Nowak}, {Oh}, {Okino}, {Olivares}, {Ortiz-Le{\'o}n},
  {Oyama}, {{\"O}zel}, {Palumbo}, {Paraschos}, {Park}, {Parsons}, {Patel}, {Pen}, {Pesce}, {Pi{\'e}tu}, {Plambeck}, {PopStefanija}, {Porth}, {P{\"o}tzl}, {Prather}, {Preciado-L{\'o}pez}, {Psaltis}, {Pu}, {Ramakrishnan}, {Rao}, {Rawlings}, {Raymond}, {Rezzolla}, {Ricarte}, {Ripperda}, {Roelofs}, {Rogers}, {Ros}, {Romero-Ca{\~n}izales}, {Roshanineshat}, {Rottmann}, {Roy}, {Ruiz}, {Ruszczyk}, {Rygl}, {S{\'a}nchez}, {S{\'a}nchez-Arg{\"u}elles}, {S{\'a}nchez-Portal}, {Sasada}, {Satapathy}, {Savolainen}, {Schloerb}, {Schonfeld}, {Schuster}, {Shao}, {Shen}, {Small}, {Sohn}, {SooHoo}, {Souccar}, {Sun}, {Tazaki}, {Tetarenko}, {Tiede}, {Tilanus}, {Titus}, {Torne}, {Traianou}, {Trent}, {Trippe}, {Turk}, {van Bemmel}, {van Langevelde}, {van Rossum}, {Vos}, {Wagner}, {Ward-Thompson}, {Wardle}, {Weintroub}, {Wex}, {Wharton}, {Wielgus}, {Wiik}, {Witzel}, {Wondrak}, {Wong}, {Wu}, {Yamaguchi}, {Yoon}, {Young}, {Young}, {Younsi}, {Yuan}, {Yuan}, {Zensus}, {Zhang}, {Zhao}, {Zhao}, {Agurto}, {Allardi}, {Amestica}, {Araneda},
  {Arriagada}, {Berghuis}, {Bertarini}, {Berthold}, {Blanchard}, {Brown}, {C{\'a}rdenas}, {Cantzler}, {Caro}, {Castillo-Dom{\'\i}nguez}, {Chan}, {Chang}, {Chang}, {Chang}, {Chang}, {Chen}, {Chilson}, {Chuter}, {Ciechanowicz}, {Colin-Beltran}, {Coulson}, {Crowley}, {Degenaar}, {Dornbusch}, {Dur{\'a}n}, {Everett}, {Faber}, {Forster}, {Fuchs}, {Gale}, {Geertsema}, {Gonz{\'a}lez}, {Graham}, {Gueth}, {Halverson}, {Han}, {Han}, {Hasegawa}, {Hern{\'a}ndez-Rebollar}, {Herrera}, {Herrero-Illana}, {Heyminck}, {Hirota}, {Hoge}, {Hostler Schimpf}, {Howie}, {Huang}, {Jiang}, {Jinchi}, {John}, {Kimura}, {Klein}, {Kubo}, {Kuroda}, {Kwon}, {Lacasse}, {Laing}, {Leitch}, {Li}, {Liu}, {Liu}, {Lin}, {Lu}, {Mac-Auliffe}, {Martin-Cocher}, {Matulonis}, {Maute}, {Messias}, {Meyer-Zhao}, {Monta{\~n}a}, {Montenegro-Montes}, {Montgomerie}, {Moreno Nolasco}, {Muders}, {Nishioka}, {Norton}, {Nystrom}, {Ogawa}, {Olivares}, {Oshiro}, {P{\'e}rez-Beaupuits}, {Parra}, {Phillips}, {Poirier}, {Pradel}, {Qiu}, {Raffin}, {Rahlin}, {Ram{\'\i}rez},
  {Ressler}, {Reynolds}, {Rodr{\'\i}guez-Montoya}, {Saez-Madain}, {Santana}, {Shaw}, {Shirkey}, {Silva}, {Snow}, {Sousa}, {Sridharan}, {Stahm}, {Stark}, {Test}, {Torstensson}, {Venegas}, {Walther}, {Wei}, {White}, {Wieching}, {Wijnands}, {Wouterloot}, {Yu}, {Yu (于威)}, \& {Zeballos}}]{2022ApJ...930L..12E}
---. 2022{\natexlab{a}}, \apjl, 930, L12

\bibitem[{{Event Horizon Telescope Collaboration} {et~al.}(2022{\natexlab{b}}){Event Horizon Telescope Collaboration}, {Akiyama}, {Alberdi}, {Alef}, {Algaba}, {Anantua}, {Asada}, {Azulay}, {Bach}, {Baczko}, {Ball}, {Balokovi{\'c}}, {Barrett}, {Baub{\"o}ck}, {Benson}, {Bintley}, {Blackburn}, {Blundell}, {Bouman}, {Bower}, {Boyce}, {Bremer}, {Brinkerink}, {Brissenden}, {Britzen}, {Broderick}, {Broguiere}, {Bronzwaer}, {Bustamante}, {Byun}, {Carlstrom}, {Ceccobello}, {Chael}, {Chan}, {Chatterjee}, {Chatterjee}, {Chen}, {Chen}, {Cheng}, {Cho}, {Christian}, {Conroy}, {Conway}, {Cordes}, {Crawford}, {Crew}, {Cruz-Osorio}, {Cui}, {Davelaar}, {De Laurentis}, {Deane}, {Dempsey}, {Desvignes}, {Dexter}, {Dhruv}, {Doeleman}, {Dougal}, {Dzib}, {Eatough}, {Emami}, {Falcke}, {Farah}, {Fish}, {Fomalont}, {Ford}, {Fraga-Encinas}, {Freeman}, {Friberg}, {Fromm}, {Fuentes}, {Galison}, {Gammie}, {Garc{\'\i}a}, {Gentaz}, {Georgiev}, {Goddi}, {Gold}, {G{\'o}mez-Ruiz}, {G{\'o}mez}, {Gu}, {Gurwell}, {Hada}, {Haggard}, {Haworth},
  {Hecht}, {Hesper}, {Heumann}, {Ho}, {Ho}, {Honma}, {Huang}, {Huang}, {Hughes}, {Ikeda}, {Impellizzeri}, {Inoue}, {Issaoun}, {James}, {Jannuzi}, {Janssen}, {Jeter}, {Jiang}, {Jim{\'e}nez-Rosales}, {Johnson}, {Jorstad}, {Joshi}, {Jung}, {Karami}, {Karuppusamy}, {Kawashima}, {Keating}, {Kettenis}, {Kim}, {Kim}, {Kim}, {Kim}, {Kino}, {Koay}, {Kocherlakota}, {Kofuji}, {Koch}, {Koyama}, {Kramer}, {Kramer}, {Krichbaum}, {Kuo}, {La Bella}, {Lauer}, {Lee}, {Lee}, {Leung}, {Levis}, {Li}, {Lico}, {Lindahl}, {Lindqvist}, {Lisakov}, {Liu}, {Liu}, {Liuzzo}, {Lo}, {Lobanov}, {Loinard}, {Lonsdale}, {Lu}, {Mao}, {Marchili}, {Markoff}, {Marrone}, {Marscher}, {Mart{\'\i}-Vidal}, {Matsushita}, {Matthews}, {Medeiros}, {Menten}, {Michalik}, {Mizuno}, {Mizuno}, {Moran}, {Moriyama}, {Moscibrodzka}, {M{\"u}ller}, {Mus}, {Musoke}, {Myserlis}, {Nadolski}, {Nagai}, {Nagar}, {Nakamura}, {Narayan}, {Narayanan}, {Natarajan}, {Nathanail}, {Fuentes}, {Neilsen}, {Neri}, {Ni}, {Noutsos}, {Nowak}, {Oh}, {Okino}, {Olivares}, {Ortiz-Le{\'o}n},
  {Oyama}, {{\"O}zel}, {Palumbo}, {Paraschos}, {Park}, {Parsons}, {Patel}, {Pen}, {Pesce}, {Pi{\'e}tu}, {Plambeck}, {PopStefanija}, {Porth}, {P{\"o}tzl}, {Prather}, {Preciado-L{\'o}pez}, {Psaltis}, {Pu}, {Ramakrishnan}, {Rao}, {Rawlings}, {Raymond}, {Rezzolla}, {Ricarte}, {Ripperda}, {Roelofs}, {Rogers}, {Ros}, {Romero-Ca{\~n}izales}, {Roshanineshat}, {Rottmann}, {Roy}, {Ruiz}, {Ruszczyk}, {Rygl}, {S{\'a}nchez}, {S{\'a}nchez-Arg{\"u}elles}, {S{\'a}nchez-Portal}, {Sasada}, {Satapathy}, {Savolainen}, {Schloerb}, {Schonfeld}, {Schuster}, {Shao}, {Shen}, {Small}, {Sohn}, {SooHoo}, {Souccar}, {Sun}, {Tazaki}, {Tetarenko}, {Tiede}, {Tilanus}, {Titus}, {Torne}, {Traianou}, {Trent}, {Trippe}, {Turk}, {van Bemmel}, {van Langevelde}, {van Rossum}, {Vos}, {Wagner}, {Ward-Thompson}, {Wardle}, {Weintroub}, {Wex}, {Wharton}, {Wielgus}, {Wiik}, {Witzel}, {Wondrak}, {Wong}, {Wu}, {Yamaguchi}, {Yoon}, {Young}, {Young}, {Younsi}, {Yuan}, {Yuan}, {Zensus}, {Zhang}, {Zhao}, {Zhao}, {Agurto}, {Araneda}, {Arriagada}, {Bertarini},
  {Berthold}, {Blanchard}, {Brown}, {C{\'a}rdenas}, {Cantzler}, {Caro}, {Chuter}, {Ciechanowicz}, {Coulson}, {Crowley}, {Degenaar}, {Dornbusch}, {Dur{\'a}n}, {Forster}, {Geertsema}, {Gonz{\'a}lez}, {Graham}, {Gueth}, {Han}, {Herrera}, {Herrero-Illana}, {Heyminck}, {Hoge}, {Huang}, {Jiang}, {John}, {Klein}, {Kubo}, {Kuroda}, {Kwon}, {Laing}, {Liu}, {Liu}, {Mac-Auliffe}, {Martin-Cocher}, {Matulonis}, {Messias}, {Meyer-Zhao}, {Montenegro-Montes}, {Montgomerie}, {Muders}, {Nishioka}, {Norton}, {Olivares}, {P{\'e}rez-Beaupuits}, {Parra}, {Poirier}, {Pradel}, {Raffin}, {Ram{\'\i}rez}, {Reynolds}, {Saez-Madain}, {Santana}, {Silva}, {Sousa}, {Stahm}, {Torstensson}, {Venegas}, {Walther}, {Wieching}, {Wijnands}, \& {Wouterloot}}]{2022ApJ...930L..13E}
---. 2022{\natexlab{b}}, \apjl, 930, L13

\bibitem[{{Event Horizon Telescope Collaboration} {et~al.}(2022{\natexlab{c}}){Event Horizon Telescope Collaboration}, {Akiyama}, {Alberdi}, {Alef}, {Algaba}, {Anantua}, {Asada}, {Azulay}, {Bach}, {Baczko}, {Ball}, {Balokovi{\'c}}, {Barrett}, {Baub{\"o}ck}, {Benson}, {Bintley}, {Blackburn}, {Blundell}, {Bouman}, {Bower}, {Boyce}, {Bremer}, {Brinkerink}, {Brissenden}, {Britzen}, {Broderick}, {Broguiere}, {Bronzwaer}, {Bustamante}, {Byun}, {Carlstrom}, {Ceccobello}, {Chael}, {Chan}, {Chatterjee}, {Chatterjee}, {Chen}, {Chen}, {Cheng}, {Cho}, {Christian}, {Conroy}, {Conway}, {Cordes}, {Crawford}, {Crew}, {Cruz-Osorio}, {Cui}, {Davelaar}, {De Laurentis}, {Deane}, {Dempsey}, {Desvignes}, {Dexter}, {Dhruv}, {Doeleman}, {Dougal}, {Dzib}, {Eatough}, {Emami}, {Falcke}, {Farah}, {Fish}, {Fomalont}, {Ford}, {Fraga-Encinas}, {Freeman}, {Friberg}, {Fromm}, {Fuentes}, {Galison}, {Gammie}, {Garc{\'\i}a}, {Gentaz}, {Georgiev}, {Goddi}, {Gold}, {G{\'o}mez-Ruiz}, {G{\'o}mez}, {Gu}, {Gurwell}, {Hada}, {Haggard}, {Haworth},
  {Hecht}, {Hesper}, {Heumann}, {Ho}, {Ho}, {Honma}, {Huang}, {Huang}, {Hughes}, {Ikeda}, {Impellizzeri}, {Inoue}, {Issaoun}, {James}, {Jannuzi}, {Janssen}, {Jeter}, {Jiang}, {Jim{\'e}nez-Rosales}, {Johnson}, {Jorstad}, {Joshi}, {Jung}, {Karami}, {Karuppusamy}, {Kawashima}, {Keating}, {Kettenis}, {Kim}, {Kim}, {Kim}, {Kim}, {Kino}, {Koay}, {Kocherlakota}, {Kofuji}, {Koch}, {Koyama}, {Kramer}, {Kramer}, {Krichbaum}, {Kuo}, {La Bella}, {Lauer}, {Lee}, {Lee}, {Leung}, {Levis}, {Li}, {Lico}, {Lindahl}, {Lindqvist}, {Lisakov}, {Liu}, {Liu}, {Liuzzo}, {Lo}, {Lobanov}, {Loinard}, {Lonsdale}, {Lu}, {Mao}, {Marchili}, {Markoff}, {Marrone}, {Marscher}, {Mart{\'\i}-Vidal}, {Matsushita}, {Matthews}, {Medeiros}, {Menten}, {Michalik}, {Mizuno}, {Mizuno}, {Moran}, {Moriyama}, {Moscibrodzka}, {M{\"u}ller}, {Mus}, {Musoke}, {Myserlis}, {Nadolski}, {Nagai}, {Nagar}, {Nakamura}, {Narayan}, {Narayanan}, {Natarajan}, {Nathanail}, {Fuentes}, {Neilsen}, {Neri}, {Ni}, {Noutsos}, {Nowak}, {Oh}, {Okino}, {Olivares}, {Ortiz-Le{\'o}n},
  {Oyama}, {{\"O}zel}, {Palumbo}, {Paraschos}, {Park}, {Parsons}, {Patel}, {Pen}, {Pesce}, {Pi{\'e}tu}, {Plambeck}, {PopStefanija}, {Porth}, {P{\"o}tzl}, {Prather}, {Preciado-L{\'o}pez}, {Psaltis}, {Pu}, {Ramakrishnan}, {Rao}, {Rawlings}, {Raymond}, {Rezzolla}, {Ricarte}, {Ripperda}, {Roelofs}, {Rogers}, {Ros}, {Romero-Ca{\~n}izales}, {Roshanineshat}, {Rottmann}, {Roy}, {Ruiz}, {Ruszczyk}, {Rygl}, {S{\'a}nchez}, {S{\'a}nchez-Arg{\"u}elles}, {S{\'a}nchez-Portal}, {Sasada}, {Satapathy}, {Savolainen}, {Schloerb}, {Schonfeld}, {Schuster}, {Shao}, {Shen}, {Small}, {Sohn}, {SooHoo}, {Souccar}, {Sun}, {Tazaki}, {Tetarenko}, {Tiede}, {Tilanus}, {Titus}, {Torne}, {Traianou}, {Trent}, {Trippe}, {Turk}, {van Bemmel}, {van Langevelde}, {van Rossum}, {Vos}, {Wagner}, {Ward-Thompson}, {Wardle}, {Weintroub}, {Wex}, {Wharton}, {Wielgus}, {Wiik}, {Witzel}, {Wondrak}, {Wong}, {Wu}, {Yamaguchi}, {Yoon}, {Young}, {Young}, {Younsi}, {Yuan}, {Yuan}, {Zensus}, {Zhang}, {Zhao}, \& {Zhao}}]{2022ApJ...930L..14E}
---. 2022{\natexlab{c}}, \apjl, 930, L14

\bibitem[{{Event Horizon Telescope Collaboration} {et~al.}(2022{\natexlab{d}}){Event Horizon Telescope Collaboration}, {Akiyama}, {Alberdi}, {Alef}, {Algaba}, {Anantua}, {Asada}, {Azulay}, {Bach}, {Baczko}, {Ball}, {Balokovi{\'c}}, {Barrett}, {Baub{\"o}ck}, {Benson}, {Bintley}, {Blackburn}, {Blundell}, {Bouman}, {Bower}, {Boyce}, {Bremer}, {Brinkerink}, {Brissenden}, {Britzen}, {Broderick}, {Broguiere}, {Bronzwaer}, {Bustamante}, {Byun}, {Carlstrom}, {Ceccobello}, {Chael}, {Chan}, {Chatterjee}, {Chatterjee}, {Chen}, {Chen}, {Cheng}, {Cho}, {Christian}, {Conroy}, {Conway}, {Cordes}, {Crawford}, {Crew}, {Cruz-Osorio}, {Cui}, {Davelaar}, {De Laurentis}, {Deane}, {Dempsey}, {Desvignes}, {Dexter}, {Dhruv}, {Doeleman}, {Dougal}, {Dzib}, {Eatough}, {Emami}, {Falcke}, {Farah}, {Fish}, {Fomalont}, {Ford}, {Fraga-Encinas}, {Freeman}, {Friberg}, {Fromm}, {Fuentes}, {Galison}, {Gammie}, {Garc{\'\i}a}, {Gentaz}, {Georgiev}, {Goddi}, {Gold}, {G{\'o}mez-Ruiz}, {G{\'o}mez}, {Gu}, {Gurwell}, {Hada}, {Haggard}, {Haworth},
  {Hecht}, {Hesper}, {Heumann}, {Ho}, {Ho}, {Honma}, {Huang}, {Huang}, {Hughes}, {Ikeda}, {Impellizzeri}, {Inoue}, {Issaoun}, {James}, {Jannuzi}, {Janssen}, {Jeter}, {Jiang}, {Jim{\'e}nez-Rosales}, {Johnson}, {Jorstad}, {Joshi}, {Jung}, {Karami}, {Karuppusamy}, {Kawashima}, {Keating}, {Kettenis}, {Kim}, {Kim}, {Kim}, {Kim}, {Kino}, {Koay}, {Kocherlakota}, {Kofuji}, {Koch}, {Koyama}, {Kramer}, {Kramer}, {Krichbaum}, {Kuo}, {La Bella}, {Lauer}, {Lee}, {Lee}, {Leung}, {Levis}, {Li}, {Lico}, {Lindahl}, {Lindqvist}, {Lisakov}, {Liu}, {Liu}, {Liuzzo}, {Lo}, {Lobanov}, {Loinard}, {Lonsdale}, {Lu}, {Mao}, {Marchili}, {Markoff}, {Marrone}, {Marscher}, {Mart{\'\i}-Vidal}, {Matsushita}, {Matthews}, {Medeiros}, {Menten}, {Michalik}, {Mizuno}, {Mizuno}, {Moran}, {Moriyama}, {Moscibrodzka}, {M{\"u}ller}, {Mus}, {Musoke}, {Myserlis}, {Nadolski}, {Nagai}, {Nagar}, {Nakamura}, {Narayan}, {Narayanan}, {Natarajan}, {Nathanail}, {Fuentes}, {Neilsen}, {Neri}, {Ni}, {Noutsos}, {Nowak}, {Oh}, {Okino}, {Olivares}, {Ortiz-Le{\'o}n},
  {Oyama}, {Palumbo}, {Paraschos}, {Park}, {Parsons}, {Patel}, {Pen}, {Pesce}, {Pi{\'e}tu}, {Plambeck}, {PopStefanija}, {Porth}, {P{\"o}tzl}, {Prather}, {Preciado-L{\'o}pez}, {Pu}, {Ramakrishnan}, {Rao}, {Rawlings}, {Raymond}, {Rezzolla}, {Ricarte}, {Ripperda}, {Roelofs}, {Rogers}, {Ros}, {Romero-Ca{\~n}izales}, {Roshanineshat}, {Rottmann}, {Roy}, {Ruiz}, {Ruszczyk}, {Rygl}, {S{\'a}nchez}, {S{\'a}nchez-Arg{\"u}elles}, {S{\'a}nchez-Portal}, {Sasada}, {Satapathy}, {Savolainen}, {Schloerb}, {Schonfeld}, {Schuster}, {Shao}, {Shen}, {Small}, {Sohn}, {SooHoo}, {Souccar}, {Sun}, {Tazaki}, {Tetarenko}, {Tiede}, {Tilanus}, {Titus}, {Torne}, {Traianou}, {Trent}, {Trippe}, {Turk}, {van Bemmel}, {van Langevelde}, {van Rossum}, {Vos}, {Wagner}, {Ward-Thompson}, {Wardle}, {Weintroub}, {Wex}, {Wharton}, {Wielgus}, {Wiik}, {Witzel}, {Wondrak}, {Wong}, {Wu}, {Yamaguchi}, {Yoon}, {Young}, {Young}, {Younsi}, {Yuan}, {Yuan}, {Zensus}, {Zhang}, {Zhao}, {Zhao}, \& {Chang}}]{2022ApJ...930L..15E}
---. 2022{\natexlab{d}}, \apjl, 930, L15

\bibitem[{{Event Horizon Telescope Collaboration} {et~al.}(2022{\natexlab{e}}){Event Horizon Telescope Collaboration}, {Akiyama}, {Alberdi}, {Alef}, {Algaba}, {Anantua}, {Asada}, {Azulay}, {Bach}, {Baczko}, {Ball}, {Balokovi{\'c}}, {Barrett}, {Baub{\"o}ck}, {Benson}, {Bintley}, {Blackburn}, {Blundell}, {Bouman}, {Bower}, {Boyce}, {Bremer}, {Brinkerink}, {Brissenden}, {Britzen}, {Broderick}, {Broguiere}, {Bronzwaer}, {Bustamante}, {Byun}, {Carlstrom}, {Ceccobello}, {Chael}, {Chan}, {Chatterjee}, {Chatterjee}, {Chen}, {Chen}, {Cheng}, {Cho}, {Christian}, {Conroy}, {Conway}, {Cordes}, {Crawford}, {Crew}, {Cruz-Osorio}, {Cui}, {Davelaar}, {De Laurentis}, {Deane}, {Dempsey}, {Desvignes}, {Dexter}, {Dhruv}, {Doeleman}, {Dougal}, {Dzib}, {Eatough}, {Emami}, {Falcke}, {Farah}, {Fish}, {Fomalont}, {Ford}, {Fraga-Encinas}, {Freeman}, {Friberg}, {Fromm}, {Fuentes}, {Galison}, {Gammie}, {Garc{\'\i}a}, {Gentaz}, {Georgiev}, {Goddi}, {Gold}, {G{\'o}mez-Ruiz}, {G{\'o}mez}, {Gu}, {Gurwell}, {Hada}, {Haggard}, {Haworth},
  {Hecht}, {Hesper}, {Heumann}, {Ho}, {Ho}, {Honma}, {Huang}, {Huang}, {Hughes}, {Ikeda}, {Violette Impellizzeri}, {Inoue}, {Issaoun}, {James}, {Jannuzi}, {Janssen}, {Jeter}, {Jiang}, {Jim{\'e}nez-Rosales}, {Johnson}, {Jorstad}, {Joshi}, {Jung}, {Karami}, {Karuppusamy}, {Kawashima}, {Keating}, {Kettenis}, {Kim}, {Kim}, {Kim}, {Kim}, {Kino}, {Koay}, {Kocherlakota}, {Kofuji}, {Koch}, {Koyama}, {Kramer}, {Kramer}, {Krichbaum}, {Kuo}, {La Bella}, {Lauer}, {Lee}, {Lee}, {Leung}, {Levis}, {Li}, {Lico}, {Lindahl}, {Lindqvist}, {Lisakov}, {Liu}, {Liu}, {Liuzzo}, {Lo}, {Lobanov}, {Loinard}, {Lonsdale}, {Lu}, {Mao}, {Marchili}, {Markoff}, {Marrone}, {Marscher}, {Mart{\'\i}-Vidal}, {Matsushita}, {Matthews}, {Medeiros}, {Menten}, {Michalik}, {Mizuno}, {Mizuno}, {Moran}, {Moriyama}, {Moscibrodzka}, {M{\"u}ller}, {Mus}, {Musoke}, {Myserlis}, {Nadolski}, {Nagai}, {Nagar}, {Nakamura}, {Narayan}, {Narayanan}, {Natarajan}, {Nathanail}, {Navarro Fuentes}, {Neilsen}, {Neri}, {Ni}, {Noutsos}, {Nowak}, {Oh}, {Okino}, {Olivares},
  {Ortiz-Le{\'o}n}, {Oyama}, {{\"O}zel}, {Palumbo}, {Filippos Paraschos}, {Park}, {Parsons}, {Patel}, {Pen}, {Pesce}, {Pi{\'e}tu}, {Plambeck}, {PopStefanija}, {Porth}, {P{\"o}tzl}, {Prather}, {Preciado-L{\'o}pez}, {Psaltis}, {Pu}, {Ramakrishnan}, {Rao}, {Rawlings}, {Raymond}, {Rezzolla}, {Ricarte}, {Ripperda}, {Roelofs}, {Rogers}, {Ros}, {Romero-Ca{\~n}izales}, {Roshanineshat}, {Rottmann}, {Roy}, {Ruiz}, {Ruszczyk}, {Rygl}, {S{\'a}nchez}, {S{\'a}nchez-Arg{\"u}elles}, {S{\'a}nchez-Portal}, {Sasada}, {Satapathy}, {Savolainen}, {Schloerb}, {Schonfeld}, {Schuster}, {Shao}, {Shen}, {Small}, {Sohn}, {SooHoo}, {Souccar}, {Sun}, {Tazaki}, {Tetarenko}, {Tiede}, {Tilanus}, {Titus}, {Torne}, {Traianou}, {Trent}, {Trippe}, {Turk}, {van Bemmel}, {van Langevelde}, {van Rossum}, {Vos}, {Wagner}, {Ward-Thompson}, {Wardle}, {Weintroub}, {Wex}, {Wharton}, {Wielgus}, {Wiik}, {Witzel}, {Wondrak}, {Wong}, {Wu}, {Yamaguchi}, {Yoon}, {Young}, {Young}, {Younsi}, {Yuan}, {Yuan}, {Zensus}, {Zhang}, {Zhao}, {Zhao}, {Chan}, {Qiu},
  {Ressler}, \& {White}}]{2022ApJ...930L..16E}
---. 2022{\natexlab{e}}, \apjl, 930, L16

\bibitem[{{Event Horizon Telescope Collaboration} {et~al.}(2022{\natexlab{f}}){Event Horizon Telescope Collaboration}, {Akiyama}, {Alberdi}, {Alef}, {Algaba}, {Anantua}, {Asada}, {Azulay}, {Bach}, {Baczko}, {Ball}, {Balokovi{\'c}}, {Barrett}, {Baub{\"o}ck}, {Benson}, {Bintley}, {Blackburn}, {Blundell}, {Bouman}, {Bower}, {Boyce}, {Bremer}, {Brinkerink}, {Brissenden}, {Britzen}, {Broderick}, {Broguiere}, {Bronzwaer}, {Bustamante}, {Byun}, {Carlstrom}, {Ceccobello}, {Chael}, {Chan}, {Chatterjee}, {Chatterjee}, {Chen}, {Chen}, {Cheng}, {Cho}, {Christian}, {Conroy}, {Conway}, {Cordes}, {Crawford}, {Crew}, {Cruz-Osorio}, {Cui}, {Davelaar}, {De Laurentis}, {Deane}, {Dempsey}, {Desvignes}, {Dexter}, {Dhruv}, {Doeleman}, {Dougal}, {Dzib}, {Eatough}, {Emami}, {Falcke}, {Farah}, {Fish}, {Fomalont}, {Ford}, {Fraga-Encinas}, {Freeman}, {Friberg}, {Fromm}, {Fuentes}, {Galison}, {Gammie}, {Garc{\'\i}a}, {Gentaz}, {Georgiev}, {Goddi}, {Gold}, {G{\'o}mez-Ruiz}, {G{\'o}mez}, {Gu}, {Gurwell}, {Hada}, {Haggard}, {Haworth},
  {Hecht}, {Hesper}, {Heumann}, {Ho}, {Ho}, {Honma}, {Huang}, {Huang}, {Hughes}, {Ikeda}, {Impellizzeri}, {Inoue}, {Issaoun}, {James}, {Jannuzi}, {Janssen}, {Jeter}, {Jiang}, {Jim{\'e}nez-Rosales}, {Johnson}, {Jorstad}, {Joshi}, {Jung}, {Karami}, {Karuppusamy}, {Kawashima}, {Keating}, {Kettenis}, {Kim}, {Kim}, {Kim}, {Kim}, {Kino}, {Koay}, {Kocherlakota}, {Kofuji}, {Koch}, {Koyama}, {Kramer}, {Kramer}, {Krichbaum}, {Kuo}, {La Bella}, {Lauer}, {Lee}, {Lee}, {Leung}, {Levis}, {Li}, {Lico}, {Lindahl}, {Lindqvist}, {Lisakov}, {Liu}, {Liu}, {Liuzzo}, {Lo}, {Lobanov}, {Loinard}, {Lonsdale}, {Lu}, {Mao}, {Marchili}, {Markoff}, {Marrone}, {Marscher}, {Mart{\'\i}-Vidal}, {Matsushita}, {Matthews}, {Medeiros}, {Menten}, {Michalik}, {Mizuno}, {Mizuno}, {Moran}, {Moriyama}, {Moscibrodzka}, {M{\"u}ller}, {Mus}, {Musoke}, {Myserlis}, {Nadolski}, {Nagai}, {Nagar}, {Nakamura}, {Narayan}, {Narayanan}, {Natarajan}, {Nathanail}, {Fuentes}, {Neilsen}, {Neri}, {Ni}, {Noutsos}, {Nowak}, {Oh}, {Okino}, {Olivares}, {Ortiz-Le{\'o}n},
  {Oyama}, {{\"O}zel}, {Palumbo}, {Paraschos}, {Park}, {Parsons}, {Patel}, {Pen}, {Pesce}, {Pi{\'e}tu}, {Plambeck}, {PopStefanija}, {Porth}, {P{\"o}tzl}, {Prather}, {Preciado-L{\'o}pez}, {Psaltis}, {Pu}, {Ramakrishnan}, {Rao}, {Rawlings}, {Raymond}, {Rezzolla}, {Ricarte}, {Ripperda}, {Roelofs}, {Rogers}, {Ros}, {Romero-Ca{\~n}izales}, {Roshanineshat}, {Rottmann}, {Roy}, {Ruiz}, {Ruszczyk}, {Rygl}, {S{\'a}nchez}, {S{\'a}nchez-Arg{\"u}elles}, {S{\'a}nchez-Portal}, {Sasada}, {Satapathy}, {Savolainen}, {Schloerb}, {Schonfeld}, {Schuster}, {Shao}, {Shen}, {Small}, {Sohn}, {SooHoo}, {Souccar}, {Sun}, {Tazaki}, {Tetarenko}, {Tiede}, {Tilanus}, {Titus}, {Torne}, {Traianou}, {Trent}, {Trippe}, {Turk}, {van Bemmel}, {van Langevelde}, {van Rossum}, {Vos}, {Wagner}, {Ward-Thompson}, {Wardle}, {Weintroub}, {Wex}, {Wharton}, {Wielgus}, {Wiik}, {Witzel}, {Wondrak}, {Wong}, {Wu}, {Yamaguchi}, {Yoon}, {Young}, {Young}, {Younsi}, {Yuan}, {Yuan}, {Zensus}, {Zhang}, {Zhao}, \& {Zhao}}]{2022ApJ...930L..17E}
---. 2022{\natexlab{f}}, \apjl, 930, L17

\bibitem[{{Falcke} {et~al.}(2000){Falcke}, {Melia}, \& {Agol}}]{2000ApJ...528L..13F}
{Falcke}, H., {Melia}, F., \& {Agol}, E. 2000, \apjl, 528, L13

\bibitem[{{Foreman-Mackey} {et~al.}(2013){Foreman-Mackey}, {Conley}, {Meierjurgen Farr}, {Hogg}, {Lang}, {Marshall}, {Price-Whelan}, {Sanders}, \& {Zuntz}}]{emcee}
{Foreman-Mackey}, D., {Conley}, A., {Meierjurgen Farr}, W., {et~al.} 2013, {emcee: The MCMC Hammer}, Astrophysics Source Code Library, record ascl:1303.002

\bibitem[{{Fox}(2017)}]{2017AGUFMSH21C..02F}
{Fox}, N.~J. 2017, in AGU Fall Meeting Abstracts, Vol. 2017, SH21C--02

\bibitem[{{Fuerst} \& {Wu}(2004)}]{2004A&A...424..733F}
{Fuerst}, S.~V., \& {Wu}, K. 2004, \aap, 424, 733

\bibitem[{{Gammie} {et~al.}(2003){Gammie}, {McKinney}, \& {T{\'o}th}}]{Gammie2013}
{Gammie}, C.~F., {McKinney}, J.~C., \& {T{\'o}th}, G. 2003, \apj, 589, 444

\bibitem[{{Georgiev} {et~al.}(2022){Georgiev}, {Pesce}, {Broderick}, {Wong}, {Dhruv}, {Wielgus}, {Gammie}, {Chan}, {Chatterjee}, {Emami}, {Mizuno}, {Gold}, {Fromm}, {Ricarte}, {Yoon}, {Joshi}, {Prather}, {Cruz-Osorio}, {Johnson}, {Porth}, {Olivares}, {Younsi}, {Rezzolla}, {Vos}, {Qiu}, {Nathanail}, {Narayan}, {Chael}, {Anantua}, {Moscibrodzka}, {Akiyama}, {Alberdi}, {Alef}, {Algaba}, {Asada}, {Azulay}, {Bach}, {Baczko}, {Ball}, {Balokovi{\'c}}, {Barrett}, {Baub{\"o}ck}, {Benson}, {Bintley}, {Blackburn}, {Blundell}, {Bouman}, {Bower}, {Boyce}, {Bremer}, {Brinkerink}, {Brissenden}, {Britzen}, {Broguiere}, {Bronzwaer}, {Bustamante}, {Byun}, {Carlstrom}, {Ceccobello}, {Chatterjee}, {Chen}, {Chen}, {Cheng}, {Cho}, {Christian}, {Conroy}, {Conway}, {Cordes}, {Crawford}, {Crew}, {Cui}, {Davelaar}, {De Laurentis}, {Deane}, {Dempsey}, {Desvignes}, {Dexter}, {Doeleman}, {Dougal}, {Dzib}, {Eatough}, {Falcke}, {Farah}, {Fish}, {Fomalont}, {Ford}, {Fraga-Encinas}, {Freeman}, {Friberg}, {Fuentes}, {Galison}, {Garc{\'\i}a},
  {Gentaz}, {Goddi}, {G{\'o}mez-Ruiz}, {G{\'o}mez}, {Gu}, {Gurwell}, {Hada}, {Haggard}, {Haworth}, {Hecht}, {Hesper}, {Heumann}, {Ho}, {Ho}, {Honma}, {Huang}, {Huang}, {Hughes}, {Ikeda}, {Impellizzeri}, {Inoue}, {Issaoun}, {James}, {Jannuzi}, {Janssen}, {Jeter}, {Jiang}, {Jim{\'e}nez-Rosales}, {Jorstad}, {Jung}, {Karami}, {Karuppusamy}, {Kawashima}, {Keating}, {Kettenis}, {Kim}, {Kim}, {Kim}, {Kim}, {Kino}, {Koay}, {Kocherlakota}, {Kofuji}, {Koch}, {Koyama}, {Kramer}, {Kramer}, {Krichbaum}, {Kuo}, {La Bella}, {Lauer}, {Lee}, {Lee}, {Lehner}, {Leung}, {Levis}, {Li}, {Lico}, {Lindahl}, {Lindqvist}, {Lisakov}, {Liu}, {Liu}, {Liuzzo}, {Lo}, {Lobanov}, {Loinard}, {Lonsdale}, {Lu}, {Mao}, {Marchili}, {Markoff}, {Marrone}, {Marscher}, {Mart{\'\i}-Vidal}, {Matsushita}, {Matthews}, {Menten}, {Michalik}, {Mizuno}, {Moran}, {Moriyama}, {M{\"u}ller}, {Mus}, {Musoke}, {Myserlis}, {Nadolski}, {Nagai}, {Nagar}, {Nakamura}, {Narayanan}, {Natarajan}, {Navarro Fuentes}, {Neilsen}, {Neri}, {Ni}, {Noutsos}, {Nowak}, {Oh},
  {Okino}, {Ortiz-Le{\'o}n}, {Oyama}, {Palumbo}, {Paraschos}, {Park}, {Parsons}, {Patel}, {Pen}, {Pi{\'e}tu}, {Plambeck}, {PopStefanija}, {P{\"o}tzl}, {Preciado-L{\'o}pez}, {Pu}, {Ramakrishnan}, {Rao}, {Rawlings}, {Raymond}, {Ripperda}, {Roelofs}, {Rogers}, {Ros}, {Romero-Ca{\~n}izales}, {Roshanineshat}, {Rottmann}, {Roy}, {Ruiz}, {Ruszczyk}, {Rygl}, {S{\'a}nchez}, {S{\'a}nchez-Arg{\"u}elles}, {S{\'a}nchez-Portal}, {Sasada}, {Satapathy}, {Savolainen}, {Schloerb}, {Schonfeld}, {Schuster}, {Shao}, {Shen}, {Small}, {Sohn}, {SooHoo}, {Souccar}, {Sun}, {Tazaki}, {Tetarenko}, {Tiede}, {Tilanus}, {Titus}, {Torne}, {Traianou}, {Trent}, {Trippe}, {Turk}, {van Bemmel}, {van Langevelde}, {van Rossum}, {Wagner}, {Ward-Thompson}, {Wardle}, {Weintroub}, {Wex}, {Wharton}, {Wiik}, {Witzel}, {Wondrak}, {Wu}, {Yamaguchi}, {Young}, {Young}, {Yuan}, {Yuan}, {Zensus}, {Zhang}, {Zhao}, \& {Zhao}}]{2022ApJ...930L..20G}
{Georgiev}, B., {Pesce}, D.~W., {Broderick}, A.~E., {et~al.} 2022, \apjl, 930, L20

\bibitem[{{Goldreich} \& {Sridhar}(1995)}]{1995ApJ...438..763G}
{Goldreich}, P., \& {Sridhar}, S. 1995, \apj, 438, 763

\bibitem[{{Igumenshchev} {et~al.}(2003){Igumenshchev}, {Narayan}, \& {Abramowicz}}]{2003ApJ...592.1042I}
{Igumenshchev}, I.~V., {Narayan}, R., \& {Abramowicz}, M.~A. 2003, \apj, 592, 1042

\bibitem[{{Iroshnikov}(1963)}]{1963AZh....40..742I}
{Iroshnikov}, P.~S. 1963, \azh, 40, 742

\bibitem[{{Kolmogorov}(1941)}]{1941DoSSR..32...16K}
{Kolmogorov}, A.~N. 1941, Akademiia Nauk SSSR Doklady, 32, 16

\bibitem[{{Kraichnan}(1965)}]{1965PhFl....8.1385K}
{Kraichnan}, R.~H. 1965, Physics of Fluids, 8, 1385

\bibitem[{{Liska} {et~al.}(2018){Liska}, {Hesp}, {Tchekhovskoy}, {Ingram}, {van der Klis}, \& {Markoff}}]{2018MNRAS.474L..81L}
{Liska}, M., {Hesp}, C., {Tchekhovskoy}, A., {et~al.} 2018, \mnras, 474, L81

\bibitem[{{Luminet}(1979)}]{1979A&A....75..228L}
{Luminet}, J.~P. 1979, \aap, 75, 228

\bibitem[{{Megale} {et~al.}(2025){Megale}, {Cruz-Osorio}, {Ficarra}, {Imbrogno}, {Meringolo}, {Primavera}, {Rezzolla}, \& {Servidio}}]{2025arXiv250904566M}
{Megale}, R., {Cruz-Osorio}, A., {Ficarra}, G., {et~al.} 2025, arXiv e-prints, arXiv:2509.04566

\bibitem[{{Mignone} {et~al.}(2007){Mignone}, {Bodo}, {Massaglia}, {Matsakos}, {Tesileanu}, {Zanni}, \& {Ferrari}}]{2007ApJS..170..228M}
{Mignone}, A., {Bodo}, G., {Massaglia}, S., {et~al.} 2007, \apjs, 170, 228

\bibitem[{{Mo{\'s}cibrodzka} \& {Gammie}(2018)}]{2018MNRAS.475...43M}
{Mo{\'s}cibrodzka}, M., \& {Gammie}, C.~F. 2018, \mnras, 475, 43

\bibitem[{{Narayan} {et~al.}(2022){Narayan}, {Chael}, {Chatterjee}, {Ricarte}, \& {Curd}}]{2022MNRAS.511.3795N}
{Narayan}, R., {Chael}, A., {Chatterjee}, K., {Ricarte}, A., \& {Curd}, B. 2022, \mnras, 511, 3795

\bibitem[{{Narayan} {et~al.}(2012){Narayan}, {S\"{a}dowski}, {Penna}, \& {Kulkarni}}]{2012MNRAS.426.3241N}
{Narayan}, R., {S\"{a}dowski}, A., {Penna}, R.~F., \& {Kulkarni}, A.~K. 2012, \mnras, 426, 3241

\bibitem[{{Narayan} \& {Yi}(1995)}]{1995ApJ...452..710N}
{Narayan}, R., \& {Yi}, I. 1995, \apj, 452, 710

\bibitem[{{Perez} {et~al.}(2021){Perez}, {Bourouaine}, {Chen}, \& {Raouafi}}]{2021A&A...650A..22P}
{Perez}, J.~C., {Bourouaine}, S., {Chen}, C. H.~K., \& {Raouafi}, N.~E. 2021, \aap, 650, A22

\bibitem[{{Porth} {et~al.}(2017){Porth}, {Olivares}, {Mizuno}, {Younsi}, {Rezzolla}, {Moscibrodzka}, {Falcke}, \& {Kramer}}]{2017ComAC...4....1P}
{Porth}, O., {Olivares}, H., {Mizuno}, Y., {et~al.} 2017, Computational Astrophysics and Cosmology, 4, 1

\bibitem[{{Psaltis} \& {Johannsen}(2012)}]{2012ApJ...745....1P}
{Psaltis}, D., \& {Johannsen}, T. 2012, \apj, 745, 1

\bibitem[{{Ricarte} {et~al.}(2022){Ricarte}, {Palumbo}, {Narayan}, {Roelofs}, \& {Emami}}]{2022ApJ...941L..12R}
{Ricarte}, A., {Palumbo}, D. C.~M., {Narayan}, R., {Roelofs}, F., \& {Emami}, R. 2022, \apjl, 941, L12

\bibitem[{{Sharma} {et~al.}(2025){Sharma}, {Medeiros}, {Wong}, {Chan}, {Halevi}, {Mullen}, \& {Stone}}]{2025ApJ...985...40S}
{Sharma}, A., {Medeiros}, L., {Wong}, G.~N., {et~al.} 2025, \apj, 985, 40

\bibitem[{{Shebalin} {et~al.}(1983){Shebalin}, {Matthaeus}, \& {Montgomery}}]{1983JPlPh..29..525S}
{Shebalin}, J.~V., {Matthaeus}, W.~H., \& {Montgomery}, D. 1983, Journal of Plasma Physics, 29, 525

\bibitem[{{S{\k{a}}dowski} {et~al.}(2014){S{\k{a}}dowski}, {Narayan}, {McKinney}, \& {Tchekhovskoy}}]{2014MNRAS.439..503S}
{S{\k{a}}dowski}, A., {Narayan}, R., {McKinney}, J.~C., \& {Tchekhovskoy}, A. 2014, \mnras, 439, 503

\bibitem[{{Taylor}(1938)}]{1938RSPSA.164..476T}
{Taylor}, G.~I. 1938, Proceedings of the Royal Society of London Series A, 164, 476

\bibitem[{{Treumann} {et~al.}(2019){Treumann}, {Baumjohann}, \& {Narita}}]{2019EP&S...71...41T}
{Treumann}, R.~A., {Baumjohann}, W., \& {Narita}, Y. 2019, Earth, Planets and Space, 71, 41

\bibitem[{{Tu} \& {Marsch}(1995)}]{1995SSRv...73....1T}
{Tu}, C.~Y., \& {Marsch}, E. 1995, \ssr, 73, 1

\bibitem[{{Velikhov}(1959)}]{1959JETP....9..995V}
{Velikhov}, E.~P. 1959, Soviet Journal of Experimental and Theoretical Physics, 9, 995

\bibitem[{{Verma}(2022)}]{2022arXiv220412790V}
{Verma}, M.~K. 2022, arXiv e-prints, arXiv:2204.12790

\bibitem[{{White}(2022)}]{2022ApJS..262...28W}
{White}, C.~J. 2022, \apjs, 262, 28

\bibitem[{{Yuan} \& {Narayan}(2014)}]{2014ARA&A..52..529Y}
{Yuan}, F., \& {Narayan}, R. 2014, \araa, 52, 529

\end{thebibliography}

\end{document}